\documentclass[twocolumn,preprintnumbers,a4paper,
superscriptaddress,floatfix,twoside]{revtex4}

\usepackage{bm}
\usepackage{epsfig}
\usepackage{graphics}
\usepackage{natbib}

\usepackage[l]{floatflt}
\usepackage{epsfig}
\usepackage{amssymb}
\usepackage{latexsym}
\usepackage{times}
\usepackage{slashed}
\usepackage{upgreek}
\usepackage{amsmath}
\usepackage{amsfonts}
\usepackage{amsbsy}
\usepackage{amscd}
\usepackage{bbm}

\newcommand{\eec}{\end{center}}
\newcommand{\bec}{\begin{center}}

\newcommand{\eem}{\end{matrix}}
\newcommand{\bem}{\begin{matrix}}
\newcommand{\eeq}{\end{equation}}
\newcommand{\beq}{\begin{equation}}
\newcommand{\ba}{\begin{array}}
\newcommand{\ea}{\end{array}}
\newcommand{\bea}{\begin{eqnarray}}
\newcommand{\eea}{\end{eqnarray}}
\newcommand{\baq}{\begin{eqnarray}}
\newcommand{\eaq}{\end{eqnarray}}
\newcommand{\beqs}{\begin{subequations}}
\newcommand{\eeqs}{\end{subequations}}
\newcommand{\bel}{\begin{align}}
\newcommand{\eal}{\end{align}}

\newcommand\eqs[2]{Eqs.~(\ref{#1}) and (\ref{#2})}

\newcommand\eqss[3]{Eqs.~(\ref{#1}), (\ref{#2}), and (\ref{#3})}

\newcommand{\ftn}{\footnotesize}

\newcommand{\TeV}{{\mbox{\rm TeV}}}
\newcommand{\MeV}{{\mbox{\rm MeV}}}
\newcommand{\GeV}{{\mbox{\rm GeV}}}

\newcommand{\EeV}{{\mbox{\rm EeV}}}
\newcommand{\PeV}{{\mbox{\rm PeV}}}

\newcommand{\etal}{{\it et al.\/}}

\def\lf{\left(}
\def\rg{\right)}
\newcommand\vev[1]{\langle {#1} \rangle}
\newcommand\vevi[1]{\langle {#1} \rangle_{\rm I}}
\newcommand{\Gr}{\ensuremath{\widetilde{G}}}

\newcommand{\Nhi}{\ensuremath{N_{\rm I\star}}}
\newcommand{\ks}{\ensuremath{k_\star}}
\newcommand{\Gsm}{\ensuremath{\mathbb{G}_{\rm SM}}}

\newcommand{\Vhi}{\ensuremath{V_{\rm I}}}
\newcommand{\vf}{\ensuremath{V_{\rm F}}}
\newcommand{\Hhi}{\ensuremath{H_{\rm I}}}
\newcommand{\Whi}{\ensuremath{W_{\rm I}}}

\newcommand{\Vhio}{\ensuremath{V_{\rm I0}}}

\newcommand{\mP}{\ensuremath{m_{\rm P}}}

\newcommand{\Mgut}{\ensuremath{M_{\rm GUT}}}
\newcommand{\Ggut}{\ensuremath{\mathbb{G}}}
\newcommand{\Gu}{\ensuremath{\mathbb{G}_{\rm GUT}}}
\newcommand{\Gfl}{\ensuremath{\mathbb{G}_{\rm 5_X}}}
\newcommand{\Glr}{\ensuremath{\mathbb{G}_{\rm LR}}}
\newcommand{\Gbl}{\ensuremath{\mathbb{G}_{B-L}}}

\newcommand{\hkp}{\ensuremath{\hat\kappa}}

\newcommand{\la}{\ensuremath{\lambda}}
\newcommand{\lm}{\ensuremath{\lambda_\mu}}
\newcommand{\aS}{\ensuremath{{\rm a}_S}}

\newcommand{\Gsn}{\ensuremath{\Gamma_{\rm I}}}
\newcommand{\msn}{\ensuremath{m_{\rm I}}}
\newcommand{\mgr}{\ensuremath{m_{3/2}}}
\newcommand{\mgri}{\ensuremath{m_{\rm I3/2}}}

\newcommand{\hd}{{\ensuremath{H_d}}}
\newcommand{\hu}{{\ensuremath{H_u}}}
\newcommand{\Nr}{\ensuremath{{\sf N}_{\rm G}}}

\newcommand{\ns}{\ensuremath{n_{\rm s}}}
\newcommand{\as}{\ensuremath{\alpha_{\rm s}}}

\newcommand{\om}{\ensuremath{\omega}}

\newcommand{\Dmax}{\ensuremath{\Delta_{\rm max*}}}
\newcommand{\Dex}{\ensuremath{\Delta_{\rm c\star}}}

\newcommand{\br}{\ensuremath{{\sf B}_{\rm h}}}

\newcommand{\Trh}{\ensuremath{T_{\rm rh}}}

\newcommand{\sg}{\ensuremath{\sigma}}
\newcommand{\sgx}{\ensuremath{\sigma_\star}}
\newcommand{\sgc}{\ensuremath{\sigma_{\rm c}}}
\newcommand{\sgm}{\ensuremath{\sigma_{\rm max}}}
\newcommand{\sgf}{\ensuremath{\sigma_{\rm f}}}
\newcommand{\ld}{\ensuremath{\lambda}}

\newcommand{\kp}{\ensuremath{\kappa}}

\def\ssb{\leavevmode\hbox{$\diagup$\kern-12pt\ftn\scshape
susy}}

\newcommand{\mss}{\ensuremath{\widetilde m}}

\newcommand{\hepth}[1]{{\ftn \tt hep-th/#1}}
\newcommand{\hepph}[1]{{\ftn\tt hep-ph/#1}}

\newcommand{\arxiv}[1]{{\ftn\tt  arXiv:#1}}

\newcommand{\Eref}[1]{Eq.~(\ref{#1})}
\newcommand{\Sref}[1]{Sec.~\ref{#1}}
\newcommand{\Fref}[1]{Fig.~\ref{#1}}
\newcommand{\Tref}[1]{Table~\ref{#1}}
\newcommand{\cref}[1]{Ref.~\cite{#1}}
\newcommand{\crefs}[1]{Refs.~\cite{#1}}

\def\th{{\theta}}
\def\ths{{\theta_S}}

\def\Ka{K\"{a}hler potential}
\def\Kaa{K\"{a}hler}
\def\Kam{K\"{a}hler manifold}
\def\Kap{K\"{a}hler potential}

\newcommand{\plk}{{\it Planck}}
\newcommand{\zm}{\ensuremath{Z_{-}}}
\newcommand{\zp}{\ensuremath{Z_{+}}}

\newcommand{\bdhh}{{\ensuremath{\normalsize I{\kern-2.9pt H}}}}

\newcommand{\phc}{\ensuremath{\Phi}}
\newcommand{\phcb}{\ensuremath{\bar\Phi}}
\newcommand{\what}{\ensuremath{\widehat}}
\newcommand{\wtilde}{\ensuremath{\widetilde}}

\newcommand{\mgro}{\ensuremath{m_{3/2}}}
\newcommand{\mz}{\ensuremath{m_{z}}}
\newcommand{\mzi}{\ensuremath{m_{{\rm I}z}}}
\newcommand{\mth}{\ensuremath{m_{\theta}}}
\newcommand{\mthi}{\ensuremath{m_{\rm I\theta}}}

\newcommand{\no}{\ensuremath{N}}

\def\bbet{{\bar\beta}}
\def\al{{\alpha}}

\def\bbet{{\bar\beta}}
\def\bz{{Z^*}}

\newcommand{\mcs}{\ensuremath{{G\mu_{\rm cs}}}}
\newcommand{\rcs}{\ensuremath{{r_{\rm cs}}}}
\newcommand{\ecs}{\ensuremath{{\epsilon_{\rm cs}}}}
\newcommand{\rms}{\ensuremath{{r_{\rm ms}}}}

\newcommand{\Ns}{\ensuremath{{N_{\rm I\star}}}}
\newcommand{\khi}{\ensuremath{K_{\rm I}}}

\newcommand{\khh}{\ensuremath{K_{\rm H}}}
\newcommand{\whi}{\ensuremath{W_{\rm H}}}
\newcommand{\wkhi}{\ensuremath{\wtilde K}}

\newcommand{\dK}{\ensuremath{K_\mu}}
\newcommand{\km}{\ensuremath{c_H}}
\newcommand{\dz}{\ensuremath{{\delta} z}}
\newcommand{\dzh}{\ensuremath{\what{\delta z}}}

\renewcommand{\Gsn}{\ensuremath{{\Gamma}_{\dz}}}
\newcommand{\Gth}{\ensuremath{{\Gamma}_{\th}}}
\newcommand{\Gh}{\ensuremath{{\Gamma}_{\tilde{h}}}}

\renewenvironment{subequations}{%
\refstepcounter{equation}%
\setcounter{parentequation}{\value{equation}}%
  \setcounter{equation}{0}
  \def\theequation{\theparentequation{\sf\small\alph{equation}}}%
  \ignorespaces
}{%
  \setcounter{equation}{\value{parentequation}}%
  \ignorespacesafterend
}

\begin{document}



\title{\bf\scshape Probing the Supersymmetry-Mass Scale With F-term Hybrid
Inflation}

\author{\scshape  G. Lazarides} 
\affiliation{School of Electrical \& Computer Engineering, Faculty
of Engineering, Aristotle University of Thessaloniki, GR-541 24
Thessaloniki, GREECE \\  {\sl e-mail address: }{\ftn\tt
glazarid@gen.auth.gr}}
\author{\scshape C. Pallis}
\affiliation{Laboratory of Physics, Faculty of Engineering,\\
Aristotle University of Thessaloniki,  GR-541 24 Thessaloniki,
GREECE \\  {\sl e-mail address: }{\ftn\tt kpallis@gen.auth.gr}}

\begin{abstract}

\noindent {\ftn \bf\scshape Abstract:} We consider F-term hybrid
inflation and supersymmetry breaking in the context of a model
which largely respects a global $U(1)$ $R$ symmetry. The K\"ahler
potential parameterizes the K\"ahler manifold with an enhanced
$U(1)\times(SU(1,1)/U(1))$ symmetry, where the scalar curvature of
the second factor is determined by the achievement of a
supersymmetry-breaking de Sitter vacuum without ugly tuning. The
magnitude of the emergent soft tadpole term for the inflaton can
be adjusted in the range $(1.2-460)~\TeV$ -- increasing with the
dimensionality of the representation of the waterfall fields -- so
that the inflationary observables are in agreement with the
observational requirements. The mass scale of the supersymmetric
partners turns out to lie in the region $(0.09-253)~\PeV$ which is
compatible with high-scale supersymmetry and the results of LHC on
the Higgs boson mass. The $\mu$ parameter can be generated by
conveniently applying the Giudice-Masiero mechanism and assures
the out-of-equilibrium decay of the $R$ saxion at a low reheat
temperature $\Trh\leq163~\GeV$.
\\ \\ {\scriptsize {\sf PACs numbers: 98.80.Cq, 12.60.Jv}
\hfill {\sl\bfseries Published in} {\sl Phys. Rev. D} {\bf 108}, no.9, 095055 (2023)}

\end{abstract}

\maketitle


\section{Introduction}

Among the various inflationary models -- for reviews see
\crefs{review,lectures} --, the simplest and most well-motivated
one is undoubtedly the \emph{F-term hybrid inflation} ({\sf\ftn
FHI}) model \cite{susyhybrid}.  It is tied to a renormalizable
superpotential uniquely determined by a global $U(1)$ $R$
symmetry, it does not require fine tuned parameters and
transplanckian inflaton values, and it can be naturally followed
by a \emph{Grand Unified Theory} ({\sf \ftn GUT}) phase transition
-- see, e.g.,  \crefs{buchbl, dvali,flipped}. In the original
implementation of FHI \cite{susyhybrid}, the slope of the
inflationary path which is needed to drive the inflaton towards
the \emph{Supersymmetric} ({\ftn\sf SUSY}) vacuum is exclusively
provided by the inclusion of \emph{radiative corrections}
({\sf\ftn  RCs}) in the tree level (classically flat) inflationary
potential. This version of FHI is considered as strongly
disfavored by the \plk\ data \cite{plin} fitted to the standard
power-law cosmological model with \emph{Cold Dark Matter}
({\ftn\sf CDM}) and a cosmological constant ($\Lambda$CDM). A more
complete treatment, though, incorporates also corrections
originating from \emph{supergravity} ({\sf \ftn SUGRA}) which
depend on the adopted \Ka\  \cite{pana, gpp, rlarge, kelar} as
well as soft SUSY-breaking terms \cite{mfhi, kaihi, sstad1,
sstad2, split, rlarge1}. Mildly tuning the parameters of the
relevant terms, we can achieve \cite{hinova} mostly hilltop  FHI
fully compatible with the data \cite{plin, plcp, gws} --
observationally acceptable implementations of FHI can also be
achieved by invoking a two-step inflationary scenario \cite{mhi}
or a specific generation \cite{dvali, muhi, nshi} of the $\mu$
term of the \emph{Minimal Supersymmetric Standard Model} ({\sf\ftn
MSSM}).

Out of the aforementioned realizations of FHI we focus here on the
``tadpole-assisted" one \cite{mfhi, kaihi} in which the suitable
inflationary potential is predominantly generated by the
cooperation of the RCs and the soft SUSY-breaking tadpole term. A
crucial ingredient for this is the specification of a convenient
SUSY-breaking scheme -- see, e.g.,
\crefs{buch1,nshi,high,stefan,davis}. Here, we extend the
formalism of FHI to encompass SUSY breaking by imposing a mildly
violated $R$ symmetry introduced in \cref{susyr}. Actually, it
acts as a junction mechanism of the (visible) \emph{inflationary
sector} ({\sf \ftn IS}) and the \emph{hidden sector} ({\sf \ftn
HS}). A first consequence of this combination is that the $R$
charge $2/\nu$ of the goldstino superfield -- which is related to
the geometry of the HS -- is constrained to values with $0<\nu<1$.
A second byproduct is that SUSY breaking is achieved not only in a
Minkowski vacuum, as in \cref{susyr}, but also in a \emph{de
Sitter} ({\ftn\sf dS}) one which allows us to control the
notorious \emph{Dark Energy} ({\ftn\sf DE}) problem by mildly
tuning a single superpotential parameter to a value of order
$10^{-12}$. A third consequence is the stabilization
\cite{buch1,high,stefan,davis} of the sgoldstino to low values
during FHI. Selecting minimal \Ka\ for the inflaton and computing
the suppressed contribution of the sgoldstino to the mass squared
of the inflaton, we show that the $\eta$-problem of FHI can be
elegantly resolved. After these arrangements, the imposition of
the inflationary requirements may restrict the magnitude of the
naturally induced tadpole term which is a function of the
inflationary scale $M$ and the dimensionality $\Nr$ of the
representation of the waterfall fields. The latter quantity
depends on the GUT gauge symmetry $\Ggut$ in which FHI is
embedded. We exemplify our proposal by considering three possible
$\Ggut$'s which correspond to the values $\Nr=1, 2,$ and $10$. The
analysis for the two latter $\Nr$ values is done for the first
time. We find that the required magnitude of the tadpole term
increases with $\Nr$.

For $\Nr=1$ the scale of formation of the $B-L$ \emph{cosmic
strings} ({\sf\ftn CSs}) fits well with the bound \cite{plcs0}
induced by the observations \cite{plcp} on the anisotropies of the
\emph{cosmic microwave background} ({\sf\ftn CMB}) radiation .
These $B-L$ CSs are rendered metastable, if the $U(1)_{B-L}$
symmetry is embedded in a GUT based on a group with higher rank
such as $SO(10)$. In such a case, the CS network decays generating
a stochastic background of gravitational waves that may interpret
\cite{nano1,kainano} the recent data from NANOGrav \cite{nano} and
other pulsar timing array experiments \cite{pta} -- see also
\cref{ligo}.

Finally, a solution to the $\mu$ problem of MSSM -- for an updated
review see \cref{mubaer} -- may be achieved by suitably applying
\cite{susyr} the Giudice-Masiero mechanism \cite{masiero, soft}.
Contrary to similar attempts \cite{muhi, nshi}, the $\mu$ term
here plays no role during FHI. This term assures \cite{baerh,
moduli, antrh, nsrh, full} the timely decay of the sgoldstino (or
$R$ saxion), which dominates the energy density of the Universe,
before the onset of the \emph{Big Bang Nucleosynthesis} ({\sf\ftn
BBN}) at cosmic temperature $(2-4)~\MeV$ \cite{nsref}. In a
portion of the parameter space with $3/4<\nu<1$ non-thermal
production of gravitinos ($\Gr$) is prohibited and so the
moduli-induced $\Gr$ problem \cite{koichi} can be easily eluded.
Finally, our model sheds light to the rather pressing problem of
the determination of the SUSY mass scale $\mss$ which remains open
to date \cite{baer} due to the lack of any SUSY signal in LHC --
for similar recent works see
Refs.~\cite{noscaleinfl,ant,kai,linde1,buch,king}. In particular,
our setting predicts $\mss$ close to the $\PeV$ scale and fits
well with high-scale SUSY and the Higgs boson mass discovered at
LHC \cite{lhc} if we assume a relatively low $\tan\beta$ and stop
mixing \cite{strumia}.


We describe below how we can interconnect the inflationary and the
SUSY-breaking sectors of our model in \Sref{asfhi}. Then, we
propose a resolution to the $\mu$ problem of MSSM in \Sref{pfhi}
and study the reheating process in \Sref{fhi}. We finally present
our results in \Sref{fhi4} confronting our model with a number of
constraints described in \Sref{fhi3}. Our conclusions are
discussed in \Sref{con}. General formulas for the SUGRA-induced
corrections to the potential of FHI are arranged in the Appendix.

\section{Linking FHI With the SUSY Breaking Sector}\label{asfhi}

As mentioned above, our model consists of two sectors: the HS
responsible for the F-term (spontaneous) SUSY breaking and the IS
responsible for FHI. In this Section we first -- in \Sref{asfhi1}
-- specify the conditions under which the coexistence of both
sectors can occur and then -- in \Sref{asfhi2} -- we investigate
the vacua of the theory. Finally, we derive the inflationary
potential in \Sref{asfhi3}.

\subsection{Set-up}\label{asfhi1}

Here we determine the particle content, the superpotential, and
the \Ka\ of our model. These ingredients are presented in
Secs.~\ref{asfhi1aa}, \ref{asfhi1a}, and \ref{asfhi1b}. Then in
Sec.~\ref{asfhi1c} we present the general structure of the SUGRA
scalar potential which governs the evolution of the HS and IS.

\subsubsection{Particle Content}\label{asfhi1aa}

As well-known, FHI can be implemented by introducing three
superfields $\bar{\Phi}$, $\Phi$, and $S$. The two first are
left-handed chiral superfields oppositely charged under a gauge
group $\Ggut$ whereas the latter is the inflaton and is a
$\Ggut$-singlet left-handed chiral superfield. Singlet under
$\Ggut$ is also the SUSY breaking (goldstino) superfield $Z$.

In this work we identify $\Ggut$ with three possible gauge groups
with different dimensionality $\Nr$ of the representations to
which $\bar{\Phi}$ and $\Phi$ belong. Namely, we consider

\paragraph{} $\Ggut=\Gbl$ with $\Gbl= \Gsm\times U(1)_{B-L}$,
where $\Gsm$ is the Standard Model gauge group. In this case
$\Phi$ and $\bar\Phi$ belong \cite{mfhi} to the $({\bf 1, 1}, 0,
-1)$ and $({\bf 1, 1},0, 1)$ representation of $\Gbl$ respectively
and so $\Nr=1$.

\paragraph{} $\Ggut=\Glr$ with $\Glr=SU(3)_{\rm C}\times
SU(2)_{\rm L} \times SU(2)_{\rm R} \times U(1)_{B-L}$. In this
case $\Phi$ and $\bar\Phi$ belong \cite{dvali, buchbl} to the
$({\bf 1, 1, 2}, -1)$ and $({\bf 1, 1, \bar 2}, 1)$ representation
of $\Glr$ respectively and so $\Nr=2$.

\paragraph{} $\Ggut=\Gfl$ with $\Gfl=SU(5)\times U(1)_X$, the
gauge group of the flipped $SU(5)$ model. In this case $\Phi$ and
$\bar\Phi$ belong \cite{flipped} to the $({\bf 10}, 1)$ and $({\bf
\overline{10}}, -1)$ representation of $\Gfl$ respectively and so
$\Nr=10$.

In the cases above, we assume that $\Ggut$ is completely broken
via the \emph{vacuum expectation values} ({\ftn\sf VEVs}) of
$\Phi$ and $\bar\Phi$ to $\Gsm$. No magnetic monopoles are
generated during this GUT transition, in contrast to the cases
where $\Ggut= SU(4)_{\rm C}\times SU(2)_{\rm L}\times SU(2)_{\rm
R}$, $SU(5)$, or $SO(10)$. The production of magnetic monopole can
be avoided, though, even in these groups if we adopt the shifted
\cite{shifted} or smooth \cite{smooth} variants of FHI.

\subsubsection{Superpotential}\label{asfhi1a}

The superpotential of our model has the form
\beq \label{Who} W=W_{\rm I}(S, \Phi, \bar\Phi) +W_{\rm H}(Z)
+W_{\rm GH}(Z, \phcb, \phc),\eeq
where the subscripts ``I'' and ``H'' stand for the IS and HS
respectively. The three parts of $W$ are specified as follows:

\setcounter{paragraph}{0}

\paragraph{} $\Whi$ is the IS part of $W$ written as
\cite{susyhybrid}
\beqs\beq \Whi = \kp S\left(\bar
\Phi\Phi-M^2\right),\label{whi}\eeq
where $\kp$ and $M$ are free parameters which may be made positive
by field redefinitions.

\paragraph{} $W_{\rm H}$ is the HS part of $W$ which reads
\cite{susyr}
\beq W_{\rm H} = m\mP^2 (Z/\mP)^\nu. \label{wh} \eeq
Here $\mP=2.4\times10^{18}~\GeV$ is the reduced Planck mass, $m$
is a positive free parameter with mass dimensions, and $\nu$ is an
exponent which may, in principle, acquire any real value if
$W_{\rm H}$ is considered as an effective superpotential valid
close to the non-zero vacuum value of $Z$. We will assume though
that the effective superpotential is such that only positive
powers of $Z$ appear.

\paragraph{} $W_{\rm GH}$ is an unavoidable term -- see below -- which
mixes $Z$ with $\bar{\Phi}$ and $\Phi$ and has the form
\beq W_{\rm GH} = -\la\mP(Z/\mP)^\nu \phcb\phc, \label{wgh}
\eeq\eeqs
with $\la$ a real coupling constant.

$W$ is fixed by imposing an $R$ symmetry under which $W$ and the
various superfields have the following $R$ characters
\beq R(W)=R(S)=2,~~R(Z)=2/\nu~~\mbox{and}~~R(\bar \Phi\Phi)=0.\eeq
As we will see below, we confine ourselves in the range
$3/4<\nu<1$. We assume that $W$ is holomorphic in $S$ and so $S$
appears with positive integer exponents $\nu_s$. Mixed terms of
the form $S^{\nu_s}Z^{\nu_z}$ must obey the $R$ symmetry and thus
\beq \nu_s+\nu_z/\nu=1~~\Rightarrow~~\nu_z=(1-\nu_s)\nu,\eeq
leading to negative values of $\nu_z$. Therefore no such mixed
terms appear in the superpotential.

\subsubsection{K\"{a}hler Potential}\label{asfhi1b}

The \Ka\ has two contributions
\beq \label{Kho} K=K_{\rm I}(S, \Phi, \bar\Phi)+K_{\rm H}(Z),\eeq
which are specified as follows:

\setcounter{paragraph}{0}

\paragraph{} $K_{\rm I}$ is the part of $K$ which depends on the fields
involved in FHI -- cf. \Eref{whi}. We adopt the simplest possible
form
\beqs\beq K_{\rm I} = |S|^2+|\Phi|^2+|\bar\Phi|^2, \\
\label{ki} \eeq
which parameterizes the $U(1)_S\times U(1)_\phc\times
U(1)_{\phcb}$ \Kam\ -- the indices here indicate the moduli which
parameterize the corresponding manifolds.

\paragraph{} $K_{\rm H}$ is the part of $K$ devoted to the HS.
We adopt the form introduced in \cref{susyr} where
\beq K_{\rm
H}=\no\mP^2\ln\lf1+\frac{|Z|^2-k^2\zm^4/\mP^2}{\no\mP^2}\rg,
\label{khi} \eeq\eeqs
with $Z_{\pm}=Z\pm Z^*$. Here, $k$ is a parameter which mildly
violates $R$ symmetry endowing $R$ axion with phenomenologically
acceptable mass. Despite the fact that there is no
string-theoretical motivation for $K_{\rm H}$, we consider it as
an interesting phenomenological option since it ensures a
vanishing potential energy density in the vacuum without tuning
for \beq \no=\frac{4\nu^2}{3-4\nu} \label{no}\eeq when $\nu$ is
confined to the following ranges
\beq
\frac34<\nu<\frac32~~\mbox{for}~~\no<0~~\mbox{and}~~
\nu<\frac34~~\mbox{for}~~\no>0.\\
\label{all} \eeq
As we will see below the same $\nu-N$ relation assists us to
obtain a dS vacuum of the whole field system with tunable
cosmological constant. Our favored $\nu$ range will finally be
$3/4<\nu<1$. This range is included in \Eref{all} for $\no<0$.
Therefore, $K_{\rm H}$ parameterizes the $\lf SU(1,1)/U(1)\rg_Z$
hyperbolic \Kam. The total $K$ in \Eref{Kho} enjoys an enhanced
symmetry for the $S$ and $Z$ fields, namely $U(1)_S\times \lf
SU(1,1)/U(1)\rg_Z$. Thanks to this symmetry, mixing terms of the
form $S^{\tilde \nu_s}Z^{*\tilde \nu_z}$ can be ignored although
they may be allowed by the $R$ symmetry for
$\tilde\nu_z=\nu\tilde\nu_s$.

\subsubsection{\boldmath SUGRA Potential}\label{asfhi1c}

Denoting the various superfields of our model as
$X^\al=S,Z,\Phi,\bar\Phi$ and employing the same symbol for their
complex scalar components, we can find the F--term (tree level)
SUGRA scalar potential $V_{\rm F}$ from $W$ in Eq.~(\ref{Who}) and
$K$ in \Eref{Kho} by applying the standard formula \cite{gref}
\beq V_{\rm F}=e^{K/\mP^2}\left(K^{\al\bbet}D_\al W D_\bbet
W^*-3{|W|^2/\mP^2}\right),\label{Vsugra} \eeq with
$K_{\al\bbet}=\partial_{X^\al}
\partial_{X^{*\bbet}}K$, $K^{\bbet\al}K_{\al\bar
\gamma}=\delta^\bbet_{\bar \gamma}$ and \beq D_\al
W=\partial_{X^\al} W + W\partial_{X^\al}K/\mP^2.\label{dW}\eeq
Thanks to the simple form of $K$ in \eqss{Kho}{ki}{khi}, the \Kaa\
metric $K_{\al\bbet}$ has diagonal form with only one non-trivial
element
\beq
\begin{split}& K_{ZZ^*}=\lf
N\mP^4-k^2\zm^4+\mP^2|Z|^2\rg^{-2}\mP^2N\times
\\&\lf\mP^6 N+12Nk^2\mP^4\zm^2+4k^4\zm^6+3k^2\mP^2\zm^2\zp^2\rg.\end{split}\eeq
The resulting $\vf$ can be written as
\beq\label{vft}\vf=e^\frac{K}{\mP^2}\Big(
|v_S|^2+|v_\phc|^2+|v_{\phcb}|^2+K_{ZZ^*}^{-1}|v_Z|^2-3|v_W|^2\Big),\eeq
where the individual contributions are
\beqs\bel v_S&=\kp(\phcb\phc-M^2) \lf|S/\mP|^2+1\rg\nonumber\\
&-S^*Z^\nu/\mP^{\nu+1}\lf m\mP-\ld\phcb\phc\rg,\label{vsgs}\\
v_\phc&=\kp S(M^2\mP^{-2}\phc^*-\phcb(|\phc|^2\mP^{-2}+1))\nonumber\\
&+Z^\nu\mP^{1-\nu}\lf\ld\phcb(|\phc/\mP|^2+1)-m\mP^{-1}\phc^* \rg,\label{vsgp}\\
v_Z&= \nu (Z/\mP)^{\nu-1}\lf m\mP-\ld\phcb\phc\rg\nonumber\\
&+N\lf Z^*\mP^{2}-4k^2\zm^3\rg\lf
N\mP^4-k^2\zm^4+|Z|^2\mP^2\rg^{-1}\nonumber\\
&\times\lf (Z/\mP)^{\nu}\lf m\mP^{2}-\ld\phcb\phc\rg +\kp S(\phcb\phc-M^2)\rg,~~~\label{vsgz}\\
v_W&= \kp S\mP^{-1}\lf\phcb\phc-M^2\rg+
Z^\nu\mP^{-\nu}(m\mP-\ld\phcb\phc).\label{vsgw}
\end{align}\eeqs
Note that $v_{\phcb}$ is obtained from $v_\phc$ by interchanging
$\phc$ with $\phcb$. Obviously, \Eref{no} was not imposed in the
formulas above.

D--term contributions to the total SUGRA scalar potential arise
only from the $\Ggut$ non-singlet fields. They take the form \beq
V_{\rm D}= \frac{g^2}{2}\lf|\phc|^2-|\phcb|^2\rg^2\label{Vd},\eeq
where $g$ is the gauge coupling constant of \Ggut. During FHI and
at the SUSY vacuum we confine ourselves along the D-flat direction
\beq |\bar\Phi|=|\Phi|, \label{inftr} \eeq which ensures that
$V_{\rm D}=0$.


\subsection{\boldmath SUSY and $\Ggut$ Breaking Vacuum}\label{asfhi2}

\begin{figure}[t]%
\epsfig{file=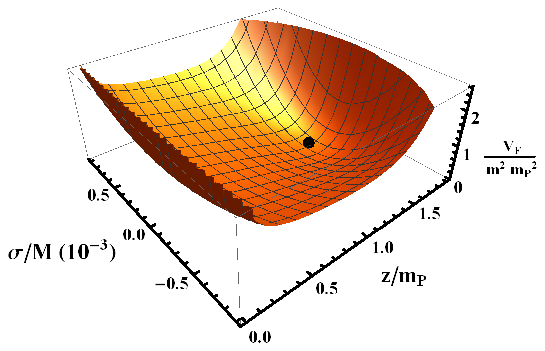,width=8.7cm,angle=-0}
\vspace*{2.3in} \caption{\sl \small The (dimensionless) SUGRA
potential $\vf/m^2\mP^2$ in \Eref{vf1vev} as a function of $z$ and
$\sg$ for the inputs shown in column B of \Tref{tab}. The location
of the dS vacuum in \Eref{zvev} is also depicted by a thick
point.} \label{fig00}
\end{figure}

As we can verify numerically, $\vf$ in \Eref{vft} is minimized at
the \Ggut-breaking vacuum
\beq \left|\vev{\Phi}\right|=\left|\vev{\bar\Phi}\right|=M.
\label{vevs} \eeq
It has also a stable valley along $\vev{\th}=0$ and
$\vev{\ths/\mP}=\pi$, with these fields defined by
\beq Z=(z+i\th)/\sqrt{2}~~\mbox{and}~~S=\sg\
e^{i\ths/\mP}/\sqrt{2}.\label{Zpara}\eeq
As we will see below, ${\ths/\mP}=\pi$ holds during FHI and we
assume that it is also valid at the vacuum. Substituting
\Eref{Zpara} in \Eref{vft}, we obtain the partially minimized
$\vf$ as a function of $z$ and $\sg$, i.e.,
\bel
&\vf(z,\sg)=2^{-(\nu+1)}e^{\vev{\khh}/\mP^2}\Bigg((\ld M^2-m\mP)^2(z/\mP)^{2\nu}\nonumber\\
&\lf
\frac{(2N\mP^2\nu+(\nu+N)z^2)^2}{N^2z^2\mP^2}-6+\frac{\sg^2}{\mP^2}\rg+\Big(2^{\frac{1+\nu}{2}}\kp
M\sg \nonumber\\
& \left.+\frac{\lf2M(\ld(M^2+\mP^2)-m\mP\rg
z^\nu}{\mP^{\nu+1}}\rg^2\Bigg).\label{vf1vev}
\end{align}
The minimization of the last term implies
\beq \label{vevsg}\sg=-2^{(1-\nu)/2}\lf\ld(M^2+\mP^2)-m\mP\rg\
z^\nu/\mP^{(\nu+1)},\eeq
whereas imposing the condition in \Eref{no} we obtain \cite{susyr}
\beq
\frac{(2N\mP^2\nu+(\nu+N)z^2)^2}{N^2z^2\mP^2}-6=\frac{(3z^2-8\nu\mP^2)^2}{16\nu^2z^2\mP^2}.
\label{nNpart}\eeq
Substituting the two last relations into \Eref{vf1vev} we arrive
at the result
\bel &\vf(z)= e^{\vev{\khh}/\mP^2}(\ld M^2-m\mP)^2 z^{2\nu}\nonumber\\
&\lf\frac{((\ld(M^2+\mP^2)-m\mP)^2}{2^{2\nu}\kp^2\mP^{4(1+\nu)}}
z^{2\nu}+\frac{(8\nu^2\mP^2-3z^2)^2}{2^{5+\nu}\nu^2z^{2}\mP^{2(\nu+1)}}\rg,
\label{vfvev}
\end{align}
which is minimized \emph{with respect to} ({\ftn\sf w.r.t.}) $\sg$
too. From the last expression we can easily find that $z$ acquires
the VEV
\beq \vev{z}=2\sqrt{2/3}|\nu|\mP, \label{zvev}\eeq
which yields the constant potential energy density
\bea \vev{\vf}=&\lf\frac{16\nu^{4}}{9}\rg^\nu \lf\frac{\ld
M^2-m\mP}{\kp\mP^2}\rg^2\om^N\times\nonumber
\\&\lf\ld(M^2+\mP^2)-m\mP\rg^2,\label{vcc}\eea
with
\beq \om=e^{\vev{\khh}/N\mP^2}\simeq2(3-2\nu)/3,\label{omg}\eeq
given that $M\ll\mP$. Tuning $\ld$ to a value $\ld\sim
m/\mP\simeq10^{-12}$ we can obtain a post-inflationary dS vacuum
which corresponds to the current DE density parameter. By virtue
of \Eref{vevsg}, we also obtain $\vev{\sg}\simeq0$.

The gravitino ($\Gr$) acquires mass \cite{gref}
\beqs\beq \label{mgr} \mgro=\vev{e^\frac{\khh}{2\mP^2}W_{\rm
H}}\simeq 2^{\nu}3^{-\nu/2} |\nu|^{\nu}m\omega^{N/2}.\eeq
Deriving the mass-squared matrix of the field system
$S-\phc-\phcb-Z$ at the vacuum we find the residual mass spectrum
of the model. Namely, we obtain a common mass for the IS
\beq \msn=e^\frac{\khh}{2\mP^2}\sqrt{2}\lf\kp^2
M^2+(4\nu^{2}/3)^\nu(1+4M^2/\mP^2)m^2\rg^{\frac12},\label{msn}\eeq
where the second term arises due to the coexistence of the IS with
the HS -- cf.~\cref{mfhi}. We also obtain the (canonically
normalized) sgoldstino (or $R$ saxion) and the pseudo-sgoldstino
(or $R$ axion) with respective masses
\beq \mz\simeq\frac{3\om}{2\nu}\mgro ~~\mbox{and}~~
\mth\simeq12k\om^{\frac32}\mgro. \label{mzth}\eeq\eeqs
Comparing the last formulas with the ones obtained in the absence
of the IS \cite{susyr} we infer that no mixing appears between the
IS and the HS. As in the ``isolated'' case of \cref{susyr} the
role of $k$ in \Eref{khi} remains crucial in providing $\th$ with
a mass. Some representative values of the masses above are
arranged in \Tref{tab} for specific $\kp, \nu,$ and $k$ values and
for the three $\Ggut$'s considered in \Sref{asfhi1aa}. We employ
values for $M$ and the tadpole parameter $\aS$ compatible with the
inflationary requirements exposed in \Sref{fhi3} -- for the
definition of $\aS$ see \Sref{asfhi3c}. We observe that $\msn$
turns out to be of order $10^{12}~\GeV$ -- cf.~\cref{mfhi} --
whereas $\mgr, \mz,$ and $\mth$ lie in the $\PeV$ range. For the
selected value $\nu=7/8>3/4$ the phenomenologically desired
hierarchy $\mz<2\mgr$ -- see \Sref{fhi3} -- is easily achieved. In
the same Table we find it convenient to accumulate the values of
some inflationary parameters introduced in Secs.~\ref{asfhi3c} and
\ref{fhi4} and some parameters related to the $\mu$ term of the
MSSM and the reheat temperature given in Secs.~\ref{pfhi} and
\ref{fhi}.

Our analytic findings related to the stabilization of the vacuum
in \eqs{vevs}{zvev} can be further confirmed by \Fref{fig00},
where the dimensionless quantity $\vf/m^2\mP^2$ in \Eref{vf1vev}
is plotted as a function of $z$ and $\sg$. We employ the values of
the parameters listed in column B of \Tref{tab}. We see that the
dS vacuum in \Eref{zvev} -- indicated by the black thick point --
is placed at $(\vev{z},\vev{\sg})=(1.43\mP,0)$ and is stable
w.r.t. both directions.

\begin{table}[t!]
\caption{\sl A Case Study Overview}
\begin{tabular}{c@{\hspace{0.3cm}}||c@{\hspace{0.9cm}}c@{\hspace{0.6cm}}c@
{\hspace{0.cm}}c} \toprule
{\sc Case:}&\hspace{0.1cm}A\hspace{0.3cm}&\hspace{0.cm}B\hspace{0.6cm}&\hspace{0.cm}C\hspace{0.cm}\\
\hline \multicolumn{4}{c}{\sc Input Parameters}\\ \hline
\multicolumn{4}{c}{$\kp=5\times10^{-4}, \nu=7/8$ ($\no=-49/8$) and $k=0.1$} \\
\hline
$\Nr$&$1$&{$2$}&$10$\\
$M~(10^{15}~\GeV)$&$1.4$&$1.9$&$3.6$\\
$m~(\PeV)$\footnote{Recall that
$1~\PeV=10^6~\GeV$ and $1~\EeV=10^9~\GeV$.}&$0.5$&$1.15$&$6.3$\\
$\ld~(10^{-12})$&$0.2$&$1.7$&$2.6$\\ \hline
\multicolumn{4}{c}{\sc HS Parameters During FHI} \\ \hline
$\vevi{z}~(10^{-3}\mP)$&$1.1$&$1.5$&$2.5$\\
$m_{\rm I3/2}~(\TeV)$&$1.2$&$2.98$&$25$\\
$m_{{\rm I}z}~(\EeV)$&0.64&$1.1$&$4.1$\\
$m_{\rm I\theta}~(\EeV)$&$0.15$&$0.32$&$1.2$\\\hline
\multicolumn{4}{c}{\sc Inflationary Parameters} \\ \hline
$\aS~(\TeV)$&$2.63$&$6.7$&$56.3$\\
$\Hhi~(\EeV)$&$0.25$&$0.4$&$1.6$\\\hline
$\sgx/\sqrt{2}M$&$1.026$&$1.035$&$1.067$\\
$\Ns$&$40.5$&$40.8$&$40.6$\\
$\Dex~(\%)$&$2.6$&$3.5$&$6.7$\\
$\Dmax~(\%)$&$2.9$&$3.9$&$7.3$\\ \hline
\multicolumn{4}{c}{\sc Inflationary Observables} \\ \hline
$\ns$&\multicolumn{3}{c}{$0.967$} \\
$-\as~(10^{-4})$&$2.3$&$2.5$&$2.9$\\
$r~(10^{-12})$&$0.9$&$3.1$&$39.7$\\\hline
\multicolumn{4}{c}{\sc Spectrum at the Vacuum} \\ \hline
$\msn~(10^{12}~\GeV)$&$1.8$&$2.4$&$4.5$\\
{$\mgro~(\PeV)$}&$0.9$&$2.$&$11.2$\\
$\mz~(\PeV)$&$1.3$&$2.9$&$16$\\
$\mth~(\PeV)$&$0.8$&$1.8$&$10$\\\hline
\multicolumn{4}{c}{\sc Reheat Temperature} \\ %
\multicolumn{4}{c}{\sc For $\mu=\mss$ ($\lm=0.69$) and $K=K_1$} \\
\hline
$\Trh~(\GeV)$&$0.07$&$0.18$&$2.05$\\ \botrule
\end{tabular}\label{tab}
\end{table}


\subsection{Inflationary Period}\label{asfhi3}

It is well known \cite{susyhybrid, lectures} that  FHI takes place
for sufficiently large $|S|$ values along a F- and D- flat
direction of the SUSY potential
\begin{equation} \label{v0}\bar\Phi={\Phi}=0, \eeq
where the potential in global SUSY
\beq \label{v00}V_{\rm SUSY}\lf{\Phi}=0\rg\equiv V_{\rm I0}=\kp^2
M^4\eeq
provides a constant potential energy density with correspoding
Hubble parameter $\Hhi=\sqrt{\Vhio/3\mP^2}$. In a SUGRA context,
though, we first check -- in Sec. \ref{asfhi3z} -- the conditions
under which such a scheme can be achieved  and then we include a
number of corrections described in Secs.~\ref{asfhi3a} and
\ref{asfhi3b} below. The final form of the inflationary potential
is given in \Sref{asfhi3c}.

\subsubsection{Hidden Sector's Stabilization}\label{asfhi3z}

\begin{figure*}[!t]
\epsfig{file=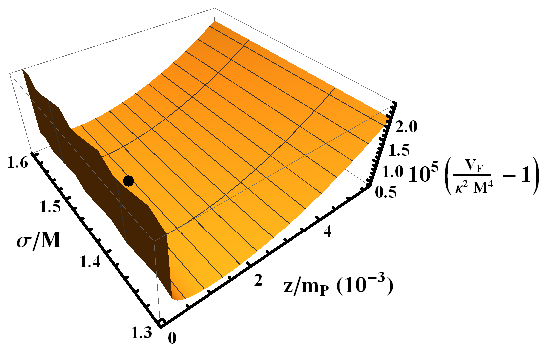,width=8.35cm,angle=-0}
\epsfig{file=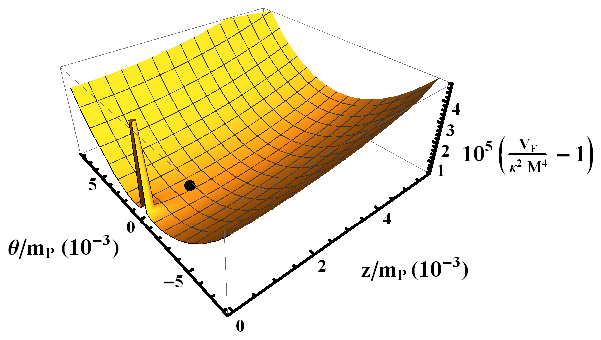,width=8.35cm,angle=-0}
\vspace*{2.3in} \caption{\sl\small The SUGRA potential
$10^5(\vf/\kp^2M^4-1)$ in \Eref{vft} along the path in \Eref{v0}
as a function of $z$ and $\sg$ for $\th=0$ (left panel) or $z$ and
$\th$ for $\sg=\sgx$ (right panel). In both cases we take the
parameters of column B in \Tref{tab}. The location of
$(\vevi{z},\sgx)$ (left panel) or $(\vevi{z},0)$ (right panel) is
also depicted by a thick black point.}\label{figVf}
\end{figure*}

The implementation of FHI is feasible in our set-up if $Z$ is well
stabilized during it. As already emphasized \cite{davis}, $V_{\rm
I0}$ in \Eref{v0} is expected to transport the value of $Z$ from
the value in \Eref{zvev} to values well below $\mP$. To determine
these values, we construct the complete expression for $\vf$ in
\Eref{vft} along the inflationary trough in \Eref{v0} and then
expand the resulting expression for low $S/\mP$ values, assuming
that the $\th=0$ direction is stable as in the vacuum. Under these
conditions $\vf$ takes the form
\beq \label{vfz}\vf(z) = e^\frac{K_{\rm H}}{\mP^2}\lf
\kp^2M^4+m^2\frac{z^{2(\nu-1)}(8\nu^2\mP^2-3z^2)^2}{2^{5+\nu}\nu^2\mP^{2\nu}}\rg.\eeq
The extremum condition obtained for $\vf(z)$ w.r.t. $z$ yields
\bel
&m^2\mP^{-2\nu}\vevi{z}^{2(\nu-2)}(64\nu^4\mP^4-9\vevi{z}^4)\times\nonumber\\&\lf8(1-\nu)\nu^2\mP^2+(3-\nu)\vevi{z}^2\rg
=2^{(7+\nu)}\nu^4\Vhio, \end{align}
where the subscript I denotes that the relevant quantity is
calculated during FHI. Given that $\vevi{z}/\mP\ll1$, the equation
above implies \beq \label{veviz}
\vevi{z}\simeq\lf\sqrt{3}\times2^{\nu/2-1}\Hhi/m\nu\sqrt{1-\nu}\rg^{1/(\nu-2)}\mP,\eeq
which is in excellent agreement with its precise numerical value.
We remark that $\nu<1$ assures the existence and the reality of
$\vevi{z}$, which is indeed much less than \mP since $\Hhi/m\ll1$.

To highlight further this key point of our scenario, we plot in
\Fref{figVf} the quantity $10^5(\vf/\kp^2M^4-1)$ with $\vf$ given
by \Eref{vft} for fixed $\phc=\phcb=0$ -- see \Eref{v0} -- and the
remaining parameters listed in column B of \Tref{tab}. In the left
panel we use as free coordinates $z$ and $\sg$ with fixed $\th=0$.
We see that the location of
$(\vevi{z},\sgx)=(1.5\times10^{-3}\mP,1.4637M)$, where $\sgx$ is
the value of $\sg$ when the pivot scale crosses outside the
horizon and is indicated by a black thick point, is independent
from $\sg$ as expected from \Eref{veviz}. In the right panel of
this figure we use as coordinates $z$ and $\th$ and fix
$\sg=\sgx$. We observe that
$(\vevi{z},\th)=(1.5\times10^{-3}\mP,0)$ -- indicated again by a
black thick point -- is well stabilized in both directions.

The (canonically normalized) components of sgoldstino acquire
masses squared, respectively,
\beqs\bea &\mzi^2\simeq6(2-\nu)\Hhi^2~~\mbox{and}~~
\mthi^2\simeq3\Hhi^2-\nonumber \eea \beq
m^2(8\nu^2\mP^2-3\vevi{z}^2)\frac{4\nu(1-\nu)\mP^2+(1-96k^2\nu)\vevi{z}^2}{2^{3+\nu}\nu\mP^{2\nu}\vevi{z}^{2(2-\nu)}},
\label{mz8i} \eeq
whereas the mass of $\Gr$ turns out to be
\beq \mgri\simeq \lf
\nu(1-\nu)^{1/2}m^{2/\nu}/\sqrt{3}\Hhi\rg^{\nu/(2-\nu)}.\eeq\eeqs
It is evident from the results above that $\mzi\gg\Hhi$ and
therefore $\vevi{z}$ is well stabilized during FHI  whereas
$m_{\rm I\th}\simeq\Hhi$ and gets slightly increased as $k$
increases. We do not think that this fact causes any problem with
isocurvature perturbations since these can be observationally
dangerous only for $\mthi\ll\Hhi$. As verified by our numerical
results, all the masses above display no $S$ dependence and so
they do not contribute to the inclination of the inflationary
potential via RCs -- see \Sref{asfhi3b} below.

\subsubsection{SUGRA Corrections}\label{asfhi3a}

The SUGRA potential in \Eref{Vsugra} induces a number of
corrections to $V_{\rm I0}$ originating not only from the IS but
also from the HS. These corrections are displayed in the Appendix
for arbitrary $\whi$ and $K_{\rm H}$. If we consider the $\whi$
and $K_{\rm H}$ in \eqs{wh}{khi} respectively, the $v$'s in
\Eref{vsgg} are found to be
\beqs\bel v_1&=2\kp M^2\mgri(2-\nu-3\vevi{z}^2/8\nu\mP^2),\label{vsg1}\\
v_2&=\kp^2 M^4 \vevi{z}^2/2\mP^2,\label{vsg2}\\
v_3&=\kp M^2 \mgri(1-\nu-3\vevi{z}^2/8\nu\mP^2),\label{vsg3}\\
v_4&=\kp^2 M^4 (1+\vevi{z}^2/\mP^2)/2.\label{vsg4}
\end{align}\eeqs
Since $\vevi{z}\ll\mP$ we do not discriminate between $\kp$ and
its rescaled form following the formulas of the Appendix. Despite
the fact that $v_2$ and $v_4$ receive contributions from both IS
and HS, as noted in the Appendix, here the IS does not participate
in $v_2$ thanks to the selected canonical \Kap\ for the $S$ field
in \Eref{ki}. This fact together with the smallness of
$\vevi{z}^2$ assists us to overcome the notorious $\eta$ problem
of FHI.

\subsubsection{Radiative Corrections}\label{asfhi3b}

These corrections originate \cite{susyhybrid} from a mass
splitting in the $\Phi-\bar{\Phi}$ supermultiplets due to SUSY
breaking on the inflationary valley. To compute them we work out
the mass spectrum of the fluctuations of the various fields about
the inflationary trough in \Eref{v0}. We obtain $2\Nr$ Weyl
fermions and $2\Nr$ pairs of real scalars with mass squared
respectively
\beq m_{\rm f}^2=\kp^2S_\ld^2
~~\mbox{and}~~m_{\pm}^2=\kp^2(S_\ld^2\pm M^2)~~\label{mscalar}\eeq
with $S_\ld=|S|-\ld\vevi{Z}^\nu\mP^{1-\nu}/\kp$. SUGRA corrections
to these masses are at most of order $M^4/\mP^2$ and can be safely
ignored. Inserting these masses into the well-known
Coleman-Weinberg formula, we find the correction
\beq V_{\rm
RC}={\kp^2\Nr\over32\pi^2}\Vhio\lf\mbox{$\sum_{i=\pm}$}
m_{i}^4\ln{m_{i}^2\over Q^2} -2m_{\rm f}^4\ln{m_{\rm f}^2 \over
Q^2}\rg,\label{Vrc}\eeq where $Q$ is a renormalization scale.
Assuming positivity of $m_{-}^2$, we obtain the lowest possible
value $S_{\rm c}$ of $S$ which assures stability of the direction
in \Eref{v0}. This critical value is equal to \beq|S_{\rm
c}|=M+\ld\vevi{Z}^\nu\mP^{1-\nu}/\kp.\label{sc}\eeq Needless to
say, the mass spectrum and $|S_{\rm c}|$ deviate slightly from
their values in the simplest model of FHI \cite{susyhybrid} due to
the mixing term in $W$ -- see \Eref{wgh}.

\subsubsection{Inflationary Potential}\label{asfhi3c}

Substituting Eqs.~(\ref{vsg1}) -- (\ref{vsg4}) into $\vf$ in
\Eref{vsgg}, including $V_{\rm RC}$ from \Eref{Vrc}, and
introducing the canonically normalized inflaton
\beq \sg=\sqrt{2K_{SS^*}}|S|~~~\mbox{with}~~ K_{SS^*}=1,
\label{sgn}\eeq the inflationary potential $V_{\rm I}$ can be cast
in the form
\beq\label{vol} V_{\rm I}\simeq V_{\rm I0}\left(1+C_{\rm
RC}+C_{\rm SSB}+C_{\rm SUGRA}\right),\eeq
where the individual contributions are specified as follows:

\setcounter{paragraph}{0}

\paragraph{} $C_{\rm RC}$ represents the RCs to
$V_{\rm I}/V_{\rm I0}$ which may be written consistently
with \Eref{Vrc} as \cite{lectures}
\beqs\beq \label{crc}C_{\rm RC}=
{\kappa^2 \Nr\over 128\pi^2}\lf8\ln{\kp^2 M^2\over Q^2}+f_{\rm
RC}(x)\rg,\eeq
with
$x=(\sigma-\sqrt{2}\ld\vevi{Z}^\nu\mP^{1-\nu}/\kp)/M>\sqrt{2}$ and
\bel f_{\rm RC}(x)&=8x^2\tanh^{-1}\lf{2}/{x^2}\rg-4(\ln4+x^4\ln
x)\nonumber \\ &+(4+x^4)\ln(x^4-4).\label{frc}\end{align}

\paragraph{} $C_{\rm SSB}$ is the contribution to $V_{\rm I}/V_{\rm I0}$
from the soft SUSY-breaking effects \cite{sstad1} parameterized as
follows:
\beq \label{cssb}C_{\rm SSB}=
m_{\rm I3/2}^2 \sg^2/2V_{\rm I0}-{\rm a}_S\,\sigma /\sqrt{2V_{\rm
I0}},\eeq
where the tadpole parameter reads
\beq \label{aSn} {\rm
a}_S=2^{1-\nu/2}m\frac{\vevi{z}^\nu}{\mP^\nu}\lf1+\frac{\vevi{z}^2}{2N\mP^2}\rg
\lf2-\nu-\frac{3\vevi{z}^2}{8\nu\mP^2}\rg.\eeq
The minus sign results from the minimization of the factor
$(S+S^*)=\sqrt{2}\sg\cos(\theta_S/\mP)$ which occurs for
$\theta_S/\mP=\pi~({\sf mod}~2\pi)$ -- the decomposition of $S$ is
shown in \Eref{Zpara}. We further assume that $\theta_S$ remains
constant during FHI. Otherwise, FHI may be analyzed as a two-field
model of inflation in the complex plane \cite{kaihi}.
Trajectories, though, far from the real axis require a significant
amount of tuning. The first term in \Eref{cssb} does not play any
essential role in our set-up due to low enough $\mgr$'s --
cf.~\cref{mfhi}.

\paragraph{} $C_{\rm SUGRA}$ is the SUGRA correction to
$V_{\rm I}/V_{\rm I0}$, after subtracting the one in $C_{\rm
SSB}$. It reads
\beq \label{csugra} C_{\rm
SUGRA}=c_{2\nu}\frac{\sg^2}{2\mP^2}+c_{4\nu}\frac{\sg^4}{4\mP^4},\eeq
where the relevant coefficients originate from \eqs{vsg2}{vsg4}
and read
\beq c_{2\nu}={\vevi{z}^2}/{2\mP^2}
~~\mbox{and}~~c_{4\nu}=(1+\vevi{z}^2/\mP^2)/2.\eeq \eeqs
Note that in similar models -- cf.~\cref{mfhi, kaihi} -- without
the presence of a HS, $c_{2\nu}$ is taken identically equal to
zero. Our present set-up shows that this assumption may be well
motivated.

\section{Generation of the $\mu$ Term of MSSM}\label{pfhi}

An important issue, usually related to the inflationary dynamics
-- see, e.g., \crefs{muhi, dvali, rsym} -- is the generation of
the $\mu$ term of MSSM. Indeed, we would like to avoid the
introduction by hand into the superpotential of MSSM of a term
$\mu\hu\hd$ with $\mu$ being an energy scale much lower than the
GUT scale -- $H_u$ and $H_d$ are the Higgs superfields coupled to
the up and down quarks respectively. To avoid this we assign $R$
charges equal to $2$ to both $H_u$ and $H_d$ whereas all the other
fields of MSSM have zero $R$ charges. Although we employ here the
notation used in a $\Gbl$ model, our construction can be easily
extended to the cases of the two other $\Ggut$'s considered -- see
\Sref{asfhi1aa}. Indeed, $H_u$ and $H_d$ are included in a
bidoublet superfield belonging to the representation $({\bf 1, 2,
2}, 0)$ in the case of $\Glr$ \cite{dvali}. On the other hand,
these superfields are included in the representations $({\bf\bar
5}, 2)$ and  $({\bf 5}, -2)$ in the case of $\Gfl$ \cite{flipped}.

The mixing term between $\hu$ and $\hd$ may emerge if we
incorporate (somehow) into the \Ka\ of our model the following
higher order terms
\beq \dK=\lm\frac{\bz^{2\nu}}{\mP^{2\nu}}\hu\hd\ +\ {\rm
h.c.},\label{dK}\eeq
where the dimensionless constant $\lm$ is taken real for
simplicity. To exemplify our approach -- cf. \cref{susyr} -- we
complement the \Ka\  in \Eref{Kho} with terms involving the
left-handed chiral superfields of MSSM denoted by $Y_\al$ with
$\al=1,...,7$, i.e.,
\bea Y_\al= {Q}, {L}, {d}^c, {u}^c, {e}^c,
\hd,~~\mbox{and}~~\hu,\nonumber \eea
where the generation indices are suppressed. Namely we consider
the following variants of the total $K$,
\beqs\bel
K_{1}&=\khh+\khi+\dK+|Y_\al|^2,~~~\label{K1}\\
K_{2}&=\no\mP^2\ln\left(1+\frac1\no\left(\frac{|Z|^2-k^2\zm^4/\mP^2}{\mP^2}+\dK\right)\rg\nonumber\\
&+\khi+|Y_\al|^2,\label{K2}\\
K_{3}&=\no\mP^2\ln\left(\frac{1+|Z|^2-k^2\zm^4/\mP^2+|Y_\al|^2}{\no\mP^2}\right)\nonumber\\
&+\khi+\dK,\label{K3}\\
K_{4}&=\no\mP^2\ln\left(1+\frac1\no\frac{|Z|^2-k^2\zm^4/\mP^2+|Y_\al|^2}{\no\mP^2}+\frac{\dK}{\no}\rg\nonumber\\
&+\khi\,.\label{K4}
\end{align}\eeqs \\ [-0.4cm]
Expanding these $K$'s for low values of $S, \phc, \phcb,$ and
$Y_\al$, we can bring them into the form
\bea K&\simeq&\khh(Z)+\khi+\wtilde
K(Z)\sum_\al|Y_\al|^2\nonumber\\&+&\lm\lf\km\frac{\bz^{2\nu}}{\mP^{2\nu}}\hu\hd+{\rm
h.c.}\rg, \label{kmssm}\eea
where $\wkhi$ is determined as follows
\beq \label{wk} \wkhi=\begin{cases} 1&\mbox{for}~~K=K_1,
K_4,\\\Big(1+\frac{|Z|^2-k^2\zm^4/\mP^2}{\mP^2\no}\Big)^{-1}&\mbox{for}~~K=K_2,
K_3,\end{cases}\eeq
whereas $\km$ is found to be
\beq \label{km} \km=\begin{cases} 1&\mbox{for}~~K=K_{1}, K_{3},\\
\Big(1+\frac{|Z|^2-k^2\zm^4/\mP^2}{\mP^2\no}\Big)^{-1}&\mbox{for}~~K=K_{2},K_{4}.\end{cases}
\eeq
Consistently with our hypothesis about the enhanced symmetry of
$K$ in \Sref{asfhi1b}, we do not consider the possibility of
including $\khi$ in the argument of the logarithm of $\khh$ as we
have done for \dK\ and/or $|Y_\al|$.

Applying the relevant formulas of \crefs{soft,susyr}, we find a
non-vanishing $\mu$ term in the superpotential of MSSM
\beq \mu \what H_u\what H_d, \eeq
where $\what Y_\al=\vev{\wtilde K}^{1/2}Y_\al$ and the $\mu$
parameter reads
\beq \label{mueff} \frac{|\mu|}{\mgr}=
\lm\lf\frac{4\nu^2}{3}\rg^\nu\times\begin{cases}
(5-4\nu)&\mbox{for}~~K=K_1,\\
3(4\nu-1)/4\nu&\mbox{for}~~K=K_{2},\\
(5-4\nu)\om&\mbox{for}~~K=K_{3},\\
3\om(4\nu-1)/4\nu&\mbox{for}~~K=K_{4}.
\end{cases}\eeq
Moreover, in the effective low energy potential we obtain a common
soft-SUSY-breaking mass parameter $\mss$ which is
\beq \label{mssi} \mss=\mgr\times\begin{cases} 1&\mbox{for}~~K=K_{1}~~\mbox{and}~~K_{2},\\
(3/2\nu-1)&\mbox{for}~~K= K_{3}~~\mbox{and}~~K_4,\end{cases}\eeq
Therefore, $\mss$ is a degenerate SUSY mass scale which can
indicatively represent the mass level of the SUSY partners. The
results in \eqs{mueff}{mssi} are consistent with those presented
in \cref{susyr}, where further details of the computation are
given.

\begin{figure}[!t]
\includegraphics[width=60mm,angle=-90]{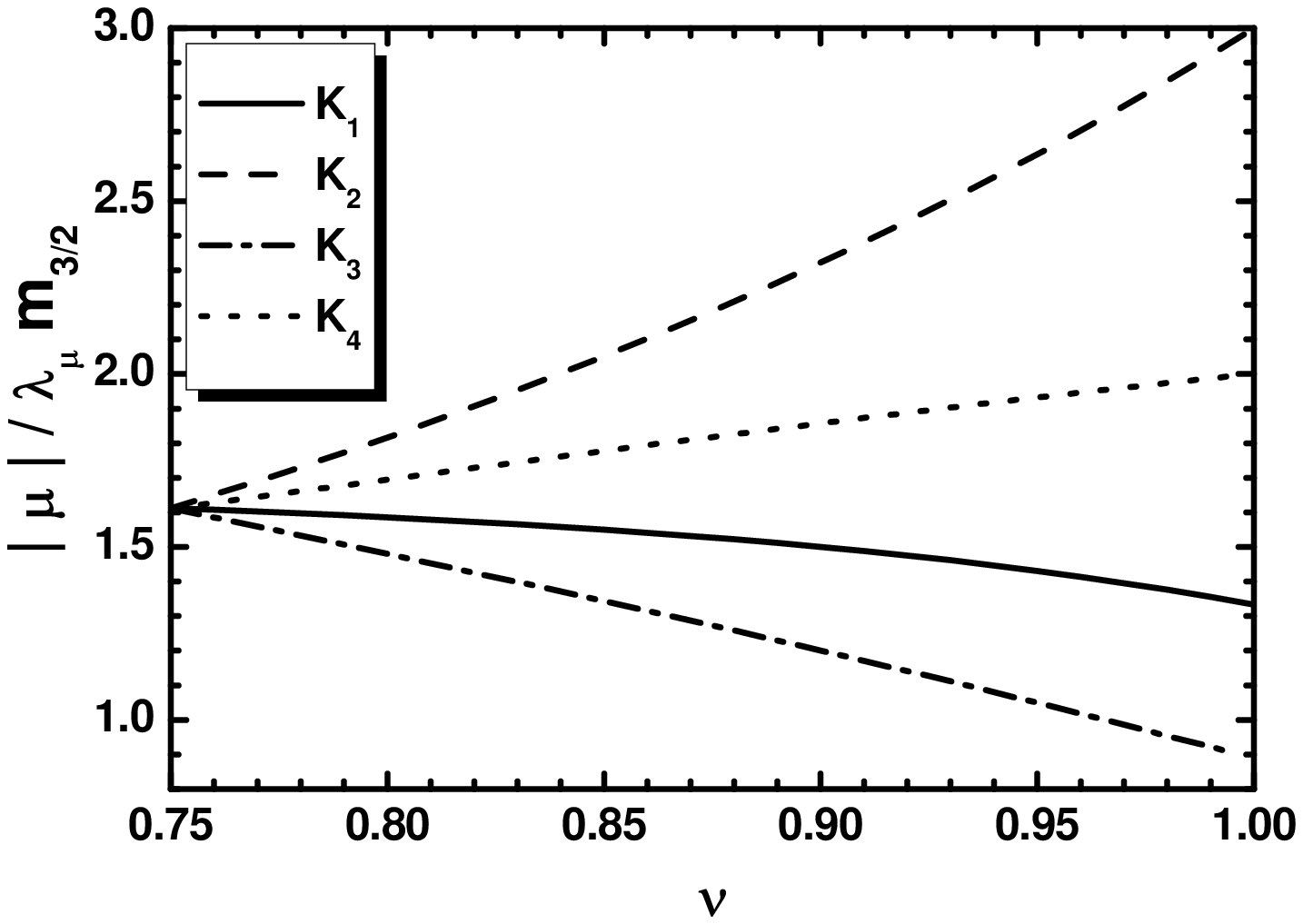}
\caption{\sl \small The ratios $|\mu|/\lm\mgr$ for $K=K_{1},
K_{2}, K_{3},$ and $K_{4}$ (solid, dashed, dot-dashed, and dotted
line respectively) versus $\nu$ in the range
$0.75-1$.}\label{fig3}
\end{figure}

The magnitude of the $\mu$'s in \Eref{mueff} is demonstrated in
\Fref{fig3}, where we present the ratios $|\mu|/\lm\mgr$ for
$K=K_{1}$ (solid line), $K_{2}$ (dashed line), $K_{3}$ (dot-dashed
line), and $K_{4}$ (dotted line) versus $\nu$ for $3/4<\nu<1$. By
coincidence all cases converge at the value
$|\mu|/\lm\mgr\simeq1.6$ for $\nu=3/4$. For $\lm$'s of order
unity, the $|\mu|$ values are a little enhanced w.r.t. $\mgr$ and
increase for $K=K_2$ and $K_4$ or decrease for $K=K_1$ and $K_3$
as $\nu$ increases.

\section{Reheating Stage}\label{fhi}

Soon after FHI the Hubble rate $H$ becomes of the order of their
masses and the IS and $z$ enter into an oscillatory phase about
their minima and eventually decay via their coupling to lighter
degrees of freedom. Note that $\th$ remains well stabilized at
zero during and after FHI and so it does not participate in the
phase of dumped oscillations. Since $\vev{z}\sim\mP$ -- see
\Eref{zvev} --, the initial energy density of its oscillations is
$\rho_{z\rm I}\sim\mz^2\vev{z}^2$. It is comparable with the
energy density of the Universe at the onset of these oscillations
$\rho_{\rm t}=3\mP^2H^2\simeq3\mP^2\mz^2$ and so we expect that
$z$ will dominate the energy density of the Universe until
completing its decay through its weak gravitational interactions.
Actually, this is a representative case of the infamous cosmic
moduli problem \cite{baerh,moduli} where reheating is induced by
long-lived massive particles with mass around the weak scale.

The reheating temperature is determined by \cite{rh}
\beq \label{Trh} \Trh= \left({72/5\pi^2g_{\rm
rh*}}\right)^{1/4}\Gsn^{1/2}\mP^{1/2},\eeq where $g_{\rm
rh*}\simeq10.75-100$ counts the effective number of the
relativistic degrees of freedom at $\Trh$. Moreover, the total
decay width $\Gsn$ of the (canonically normalized) sgoldstino
\beq\dzh=\vev{K^{1/2}_{ZZ^*}}\dz~~\mbox{with}~~\dz=z-\vev{z}~~\mbox{and}~~\vev{K_{ZZ^*}}=\vev{\om}^{-2}
\label{dphi}\eeq
predominantly includes the contributions from its decay into
pseudo-sgoldstinos and Higgs bosons via the kinetic terms
$K_{XX^*}\partial_\mu X\partial^\mu X^*$ with $X=Z, \hu$ and $\hd$
\cite{full,baerh,antrh,nsrh} of the Lagrangian. In particular, we
have
\beq\Gsn\simeq\Gth+\Gh,\label{Gol}\eeq
where the individual decay widths  are given by
\beqs\beq \Gth\simeq\frac{\ld_\th^2\mz^3}{32\pi
\mP^2}\sqrt{1-{4\mth^2}/{\mgr^2}}\label{Gth}\eeq
with $\ld_\th=-\vev{z}/{N\mP}=({4\nu-3})/{\sqrt{6}\nu}$, and
\beq \Gh=\frac{3^{2\nu+1}}{2^{4\nu+1}}\lm^2\frac{\om^2}{4\pi}
\frac{\mz^3}{\mP^2}\nu^{-4\nu}\,.\label{Gh}\eeq\eeqs
Other possible decay channels into gauge bosons through anomalies
and three-body MSSM (s)particles  are subdominant. On the other
hand, we kinematically block the decay of $\dzh$ into $\Gr$'s
\cite{koichi,baerh} in order to protect our setting from
complications with BBN due to possible late decay of the produced
$\Gr$ and problems with the abundance of the subsequently produced
lightest SUSY particles. In view of \eqs{mzth}{mgr}, this aim can
be elegantly achieved if we set $\nu>3/4$.

\begin{figure}[!t]
\includegraphics[width=60mm,angle=-90]{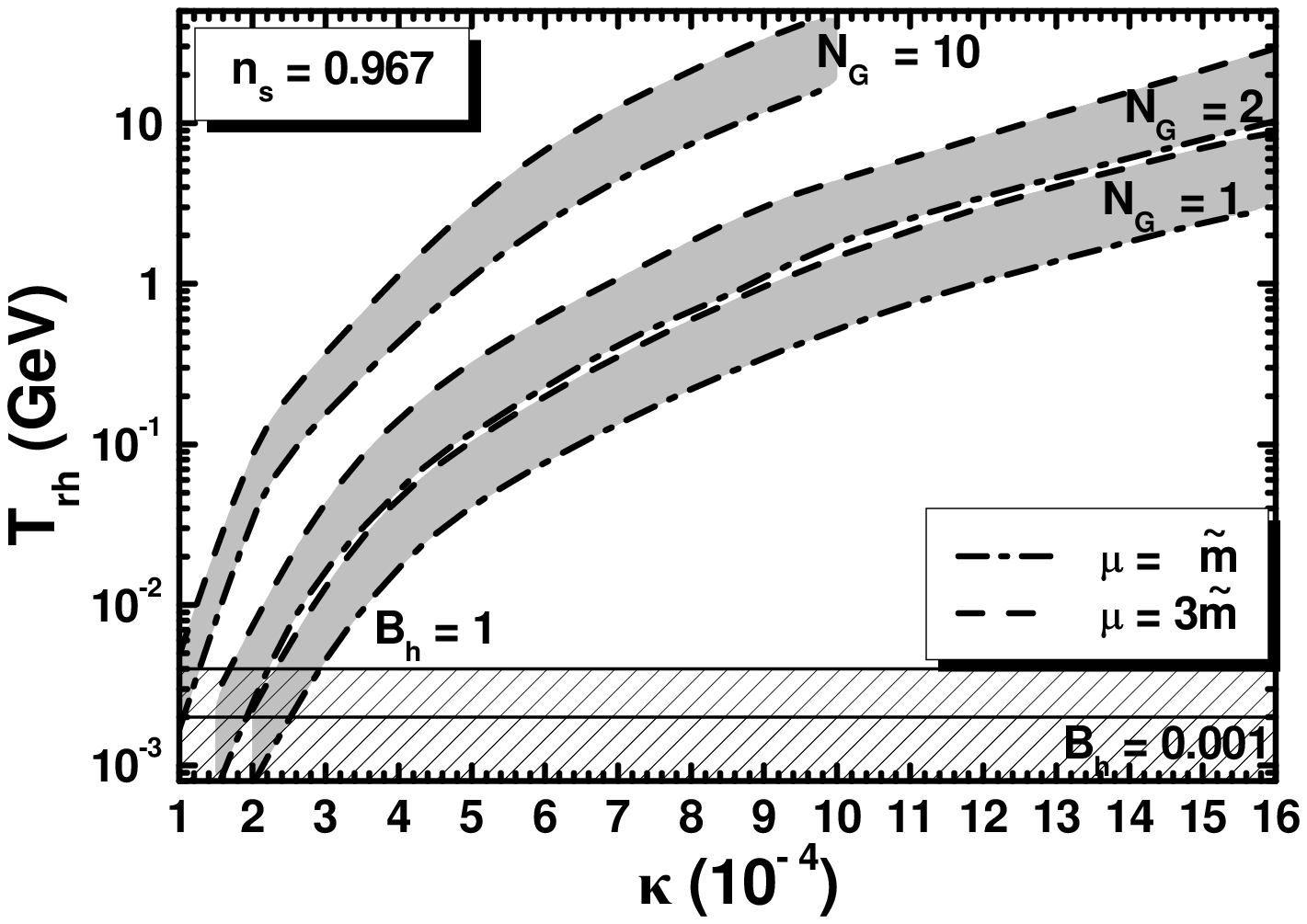}
\caption{\sl Allowed strips in the $\kp-\Trh$ plane compatible
with the inflationary requirements in \Sref{fhi3} for $\ns=0.967$,
and various $\Nr$ values indicated in the graph. We take $K=K_1$,
$\nu=7/8$, and $\mu=\mss$ (dot-dashed lines) or $\mu=3\mss$
(dashed lines). The BBN lower bounds on $\Trh$ for hadronic
branching ratios $\br=1$ and $0.001$ are also depicted by two thin
lines.}\label{figrh}
\end{figure}

Taking $\kp$ and $\mz$ values allowed by the inflationary part of
our model -- see \Sref{fhi4} -- and selecting some specific $K$
from Eqs.~(\ref{K1}) -- (\ref{K4}), we evaluate $\Trh$ as a
function of $\kp$ and determine the regions allowed by the BBN
constraints in \eqs{tns1}{tns2} -- see \Sref{fhi3} below. The
results of this computation are displayed in \Fref{figrh}, where
we design allowed contours in the $\kp-\Trh$ plane for the various
$\Nr$'s and $\nu=7/8$. This is an intermediate value in the
selected margin $(3/4-1)$. The boundary curves of the allowed
regions correspond to $\mu=\mss$ or $\lm=0.65$ (dot-dashed line)
and $\mu=3\mss$ or $\lm=1.96$ (dashed line). The $|\mu|/\mss-\lm$
correspondence is determined via \Eref{mueff} for a selected $K$.
Here we set $K=K_1$. Qualitatively similar results are obtained
for an alternative $K$ choice. We see that there is an ample
parameter space consistent with the BBN bounds depicted by two
horizontal lines. Since the satisfaction of the inflationary
requirements leads to an increase of the scale $m$ with $\Nr$ and
$m$ heavily influences $\mz$ and consequently $\Trh$ -- see
\Eref{Trh} -- this temperature increases with $\Nr$. The maximal
values of $\Trh$ for the selected $\nu$ are obtained for
$\mu=3\mss$ and are estimated to be
\beq T_{\rm rh}^{\max}\simeq14~\GeV,~33~\GeV,~\mbox{and}~~49~\GeV
\eeq
for $\Nr=1,~2,$ and $10$ respectively. Obviously, reducing $\mu$
below $\mss$, the parameters $\lm$, $\Gsn$, and so $\Trh$ decrease
too and the slice cut by the BBN bound increases. Therefore, our
setting fits better with high-scale SUSY \cite{strumia} and not
with split \cite{strumia} or natural \cite{baerh} SUSY which
assume $\mu\ll\mss$.



\section{Observational Requirements}\label{fhi3}

Our set-up must satisfy a number of observational requirements
specified below.

\setcounter{paragraph}{0}

\paragraph{} The number of e-foldings that the pivot scale
$\ks=0.05/{\rm Mpc}$ undergoes during FHI must be adequately large
for the resolution of the horizon and flatness problems of
standard Big Bang cosmology. Assuming that FHI is followed, in
turn, by a decaying-particle, radiation and matter dominated era,
we can derive the relevant condition \cite{hinova, plin}:
\begin{equation}  \label{Nhi}
\Ns=\int_{\sigma_{\rm f}}^{\sigma_{\star}} \frac{d\sigma}{m^2_{\rm
P}}\: \frac{V_{\rm I}}{V'_{\rm I}}\simeq19.4+{2\over
3}\ln{V^{1/4}_{\rm I0}\over{1~{\rm GeV}}}+ {1\over3}\ln {T_{\rm
rh}\over{1~{\rm GeV}}},
\end{equation}
where the prime denotes derivation w.r.t. $\sigma$, $\sgx$ is the
value of $\sigma$ when $\ks$ crosses outside the inflationary
horizon, and $\sigma_{\rm f}$ is the value of $\sigma$ at the end
of FHI. The latter coincides with either the critical point
$\sgc=\sqrt{2}|S_{\rm c}|$ -- see \Eref{mscalar} --, or the value
of $\sg$ for which one of the slow-roll parameters \cite{review}
\beq \label{slow} \epsilon={m^2_{\rm P}}\left({V'_{\rm
I}}/\sqrt{2}{V_{\rm I}}\right)^2~~\mbox{or}~~\eta= m^2_{\rm
P}~{V''_{\rm I}}/{V_{\rm I}} \eeq
exceeds unity in absolute value. For $\ld\sim10^{-12}$ as required
by the cosmic coincidence problem -- see below -- we obtain
$\vev{\sg}\simeq0$ which does not disturb the inflationary
dynamics since $\vev{\sg}\ll\sgc$.

\paragraph{} The amplitude $A_{\rm s}$ of the power spectrum of
the curvature perturbation generated by $\sigma$ during FHI and
calculated at $k_{\star}$ as a function of $\sgx$ must be
consistent with the data \cite{plcp}, i.e.,
\begin{equation} \label{Prob}
\sqrt{A_{\rm s}}= \frac{1}{2\sqrt{3}\, \pi m^3_{\rm P}}\;
\left.\frac{V_{\rm I}^{3/2}(\sigma_\star)}{|V'_{\rm
I}(\sigma_\star)|}\right.\simeq\: 4.588\times 10^{-5}.
\end{equation}
The observed curvature perturbation is generated wholly by
$\sigma$ since the other scalars are adequately massive during FHI
-- see \Sref{asfhi3z}.

\paragraph{} The scalar spectral index $\ns$, its running
$\as$, and the scalar-to-tensor ratio $r$ must be in agreement
with the fitting of the \plk\ TT, TE, EE+lowE+lensing, {\sl
BICEP/Keck Array} ({\sffamily\ftn BK18}), and BAO data
\cite{plin,gws} with the $\Lambda$CDM$+r$ model which
approximately requires that, at 95$\%$ \emph{confidence level}
({\sf\ftn c.l.}),
\begin{equation}  \label{nswmap}
\ns=0.967\pm0.0074\>\>\>\mbox{and}\>\>\>r\leq0.032,
\end{equation}
with $|\as|\ll0.01$. These observables are calculated employing
the standard formulas
\beqs\bel \label{nS} & \ns=1-6\epsilon_\star\ +\ 2\eta_\star,\\
& \label{aS}
\as={2}\left(4\eta_\star^2-(\ns-1)^2\right)/3-2\xi_\star,~\mbox{and}~~
r=16\epsilon_\star,  \end{align}\eeqs
where $\xi\simeq m_{\rm P}^4~V'_{\rm I} V'''_{\rm I}/V^2_{\rm I}$
and all the variables with the subscript $\star$ are evaluated at
$\sigma=\sgx$.

\paragraph{} The dimensionless tension $\mcs$ of the $B-L$ CSs produced
at the end of FHI in the case $\Ggut=\Gbl$ is \cite{mark}
\begin{equation} \label{mucs} \mcs \simeq
\frac12\lf\frac{M}{\mP}\rg^2\ecs(\rcs)~~\mbox{with}~~\ecs(\rcs)=\frac{2.4}{\ln(2/\rcs)}\cdot\end{equation}
Here $G=1/8\pi\mP^2$ is the Newton gravitational constant and
$\rcs=\kappa^2/2g^2\leq10^{-2}$ with $g\simeq0.7$ being the gauge
coupling constant at a scale close to $M$. $\mcs$ is restricted by
the level of the CS contribution to the observed anisotropies of
CMB radiation reported by \plk\ \cite{plcs0} as follows:
\beqs\beq \mcs\lesssim 2.4\times 10^{-7}~~\mbox{at 95$\%$ c.l.}
\label{plcs} \eeq
On the other hand, the primordial CS loops and segments connecting
monopole pairs decay by emitting stochastic gravitational
radiation which is measured by the pulsar timing array experiments
\cite{nano,pta}. If the CS network is stable, the recent
observations require \cite{nano1}
\beq \mcs\lesssim 2\times 10^{-10}~~ \mbox{at 95$\%$ c.l.}
\label{ppta} \eeq
However, if the CSs are metastable, due to the embedding of $\Gbl$
into a larger group $\Gu$ whose breaking leads to monopoles which
can break the CSs, the interpretation \cite{nano1} of the recent
observations \cite{nano,pta} dictates
\beq  10^{-8}\lesssim  \mcs\lesssim 2\times
10^{-7}~~\mbox{for}~~8.2\gtrsim\sqrt{\rms}\gtrsim7.9\label{kai}
\eeq\eeqs
at $2\sigma$ where the upper bound originates from \cref{ligo} and
is valid for a standard cosmological evolution and CSs produced
after inflation. Here $\rms$ is the ratio of the monopole mass
squared to $\mu_{\rm cs}$. Since we do not specify further this
possibility in our work, the last restriction does not impact on
our parameters.

\paragraph{} Consistency between theoretical and observational values of
light element abundances predicted by BBN imposes a lower bound on
$\Trh$, which depends on the mass of the decaying particle $z$ and
the hadronic branching ratio $\br$. Namely, for large
$\mz\sim10^5~\GeV$, the most up-to-date analysis of \cref{nsref}
entails
\beqs \bel&\Trh\geq4.1~\MeV~~\mbox{for}~~\br=1\label{tns1} \\
\mbox{and}~~&\Trh\geq2.1~\MeV~~\mbox{for}~~\br=10^{-3}.\label{tns2}\end{align}\eeqs
The BBN bound is mildly softened for larger $\mz$ values.
Moreover, the possible production of $\Gr$ from the $z$ decay is
mostly problematic \cite{baerh} since it may lead to
overproduction of the LSP (i.e., the lightest SUSY particle),
whose non-thermally produced abundance from the $\Gr$ decay can
drastically overshadow its thermally-produced one. As a
consequence, the LSP abundance can easily violate the
observational upper bound \cite{plcp} from CDM considerations.
This is the moduli-induced \cite{koichi} LSP overproduction
problem via the $\Gr$ decay \cite{baerh}. To avoid this
complication, we kinematically forbid the decay of $z$ into $\Gr$
selecting $\nu>3/4$ which ensures that $\mz<2\mgr$ -- see
\Eref{mzth}.

\paragraph{} We identify $\vev{\vf}$ in \Eref{vcc} with the DE energy density,
i.e.,
\beq \label{omde} \vev{\vf}=\Omega_\Lambda\rho_{\rm
c0}=7.2\times10^{-121}\mP^4,\eeq
where $\Omega_\Lambda=0.6889$ and $\rho_{\rm
c0}=2.4\times10^{-120}h^2\mP^4$ with $h=0.6732$ \cite{plcp} are
the density parameter of DE and the current critical energy
density of the Universe respectively. By virtue of \Eref{vcc}, we
see that \Eref{omde} can be satisfied for $\ld\sim m/\mP$.
Explicit values are given for the cases in \Tref{tab}.

\paragraph{} Scenarios with large $\mss$, although not directly accessible at
the LHC, can be probed via the measured value of the Higgs boson
mass. Within high-scale SUSY, updated analysis requires
\cite{lhc,strumia}
\beq
3\times10^3\lesssim\mss/\GeV\lesssim3\times10^{11},\label{highb}
\eeq
for degenerate sparticle spectrum, $\mu$ and $\tan\beta$ in the
ranges $\mss/3\leq\mu\leq3\mss$ and $1\leq\tan\beta\leq50$, and
varying the stop mixing.

\section{Results}\label{fhi4}

As deduced from Secs.~\ref{asfhi1aa} -- \ref{asfhi1b} and
\ref{pfhi}, our model depends on the parameters
\bea \Nr,~\kappa,~M,~m,~\ld,~\nu,~k,~~\mbox{and}~~\lm\nonumber
\eea
(recall that $N$ is related to $\nu$ via \Eref{no}). Let us
initially clarify that $\ld$ can be fixed at a rather low value as
explained below \Eref{omde} and does not influence the rest of our
results. Moreover, $k$ affects $\mth$ and $\mthi$ via
\eqs{mzth}{mz8i} and helps us to avoid massless modes. We take
$k=0.1$ throughout our investigation.

As shown in \cref{mfhi}, the confrontation of FHI with data for
any fixed $\Nr$ requires a specific adjustment between $\kp$ or
$M$ and the $\aS$ which is given in \Eref{aSn} as a function of
$m$, $\nu$, $\kp$, and $M$ -- see \Eref{veviz}. Obviously a
specific $\aS$ value can be obtained by several choices of the
initial parameters $\nu$ and $m$. These parameters influence also
the requirement in \Eref{Nhi} via $\Trh$, which is given in
\Eref{Trh}. However, to avoid redundant solutions we first explore
our results for the IS in terms of the variables $\kp, M,$ and
$\aS$ in \Sref{fhi4a} taking a representative $\Trh$ value, e.g.,
$T_{\rm rh}\simeq1~\GeV$. Variation of $\Trh$ over one or two
orders of magnitude does not affect our findings in any essential
way. Therefore, we do not impose in \Sref{fhi4a} the constraints
from the BBN in \eqs{tns1}{tns2}. In \Sref{fhi4b}, we then
interconnect these results with the HS parameters $\nu$ and $m$.

\subsection{Inflation Analysis}\label{fhi4a}

Enforcing  the constraints in Eqs.~(\ref{Nhi}) and (\ref{Prob}) we
can find $M$ and $\sgx$, for any given $\Nr$, as functions of our
free parameters $\kappa$ and $\aS$. Let us clarify here that for
$\Nr=1$ the parameter space is identical with the one explored in
\cref{mfhi}, where the HS is not specified. As explained there --
see also \cref{kaihi} -- observationally acceptable values of
$\ns$ can be achieved by implementing hilltop FHI. This type of
FHI requires a non-monotonic $\Vhi$ with $\sigma$ rolling from its
value $\sgm$ at which the maximum of $\Vhi$ lies down to smaller
values. As for any model of hilltop inflation, $\Vhi'$ and
therefore $\epsilon$ in \Eref{slow} and $r$ in \Eref{aS} decrease
sharply as $\Nhi$ increases -- see \Eref{Nhi} --, whereas $\Vhi''$
(or $\eta$) becomes adequately negative, thereby lowering $\ns$
within its range in \Eref{nswmap}.

These qualitative features are verified by the approximate
computation of the quantities in \Eref{slow} for $\sg<\sgm$
which are found to be
\beq\label{sr} \epsilon\simeq \mP^2\lf C'_{\rm RC}+C'_{\rm
SSB}\rg^2/2~~\mbox{and}~~ \eta\simeq \mP^2 C''_{\rm RC},\eeq
where the derivatives of the various contributions read
\beqs\bel\label{cdev1} C'_{\rm SSB}&\simeq-\aS/\sqrt{2\Vhio},\\
\label{cdev2} C'_{\rm
RC}&\simeq\frac{\Nr\kp^2x}{32M\pi^2}\lf4\tanh^{-1}\lf{2}/{x^2}
\rg+x^2\ln(1-4/x^4)\rg,\\
\label{cdev3} C''_{\rm
RC}&\simeq\frac{\Nr\kp^2}{32M^2\pi^2}\lf4\tanh^{-1}
\lf{2}/{x^2}\rg+3x^2\ln(1-4/x^4)\rg.
\end{align}\eeqs
The required behavior of $\Vhi$ in \Eref{vol} can be attained, for
given $\Nr$, thanks to the similar magnitudes and the opposite
signs of the terms $C'_{\rm RC}$ and $C'_{\rm SSB}$ in
\eqs{cdev1}{cdev2} which we can obtain for carefully selecting
$\kp$ and $\aS$. Apparently, we have $C'_{\rm SSB}<0$ and $C'_{\rm
RC}>0$ for $\sgx<\sgm$ since
$|4\tanh^{-1}\lf{2}/{x^2}\rg|>|x^2\ln(1-4/x^4)|$. On the contrary,
$C''_{\rm RC}<0$, since the negative contribution
$3x^2\ln(1-4/x^4)$ dominates over the first positive one, and so
we obtain $\eta<0$ giving rise to acceptably low $\ns$ values.

\begin{figure}[!t]
\centering\includegraphics[width=60mm,angle=-90]{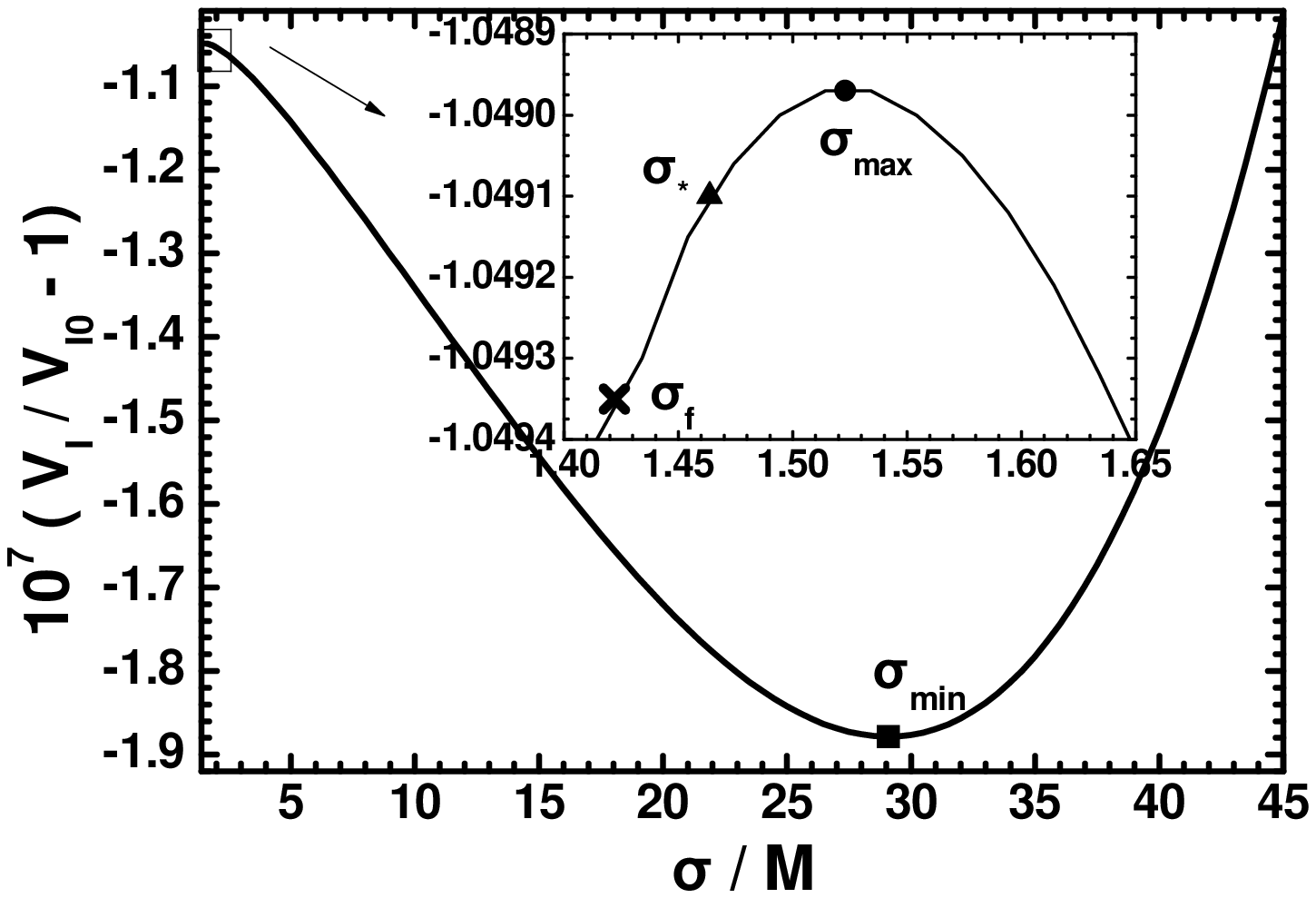}
\caption{\sl $\Vhi$ as a function of $\sg$ in units of $M$ for the
parameters given in column B of \Tref{tab}. The values $\sgx$,
$\sg_{\rm f}$, $\sg_{\rm max}$, and $\sg_{\rm min}$ of $\sigma$
are also depicted.}\label{fig0}
\end{figure}

We can roughly determine $\sgm$ by expanding $C'_{\rm RC}$ for
large $\sg$ and equating the result with $C'_{\rm SSB}$. We
obtain
\begin{equation}
\label{sgm} \frac{\Nr\kp^2}{8\pi^2\sgm}=\frac{\aS}{\sqrt{2}\kp
M^2}~\Rightarrow~\sgm\simeq
\frac{\kp^3M^2\Nr}{4\sqrt{2}\pi^2\aS}.\eeq
Needless to say, $\Vhi$ turns out to be bounded from below for
large $\sg$'s since in this regime $C_{\rm SUGRA}$ starts
dominating over $C_{\rm RC}$ generating thereby a
($\Nr$-independent) minimum at about
\begin{equation}
\label{sgmn} \sigma_{\rm min}\simeq
\lf\frac{\aS\mP^4}{\sqrt{2}c_{4\nu}\kp M^2}\rg^{1/3}\cdot\eeq
For $\sg>\sg_{\rm min}$, $\Vhi$ becomes a monotonically
increasing function of $\sg$ and so the boundedness of $\Vhi$
is assured.

From our numerical computation we observe that, for constant $\Nr,
\kp$, and $\aS$, the lower the value for $n_{\rm s}$ we wish to
attain, the closer we must set $\sgx$ to $\sgm$. Given that $\sgm$
turns out to be comparable to $\sgc$ and the hierarchy
$\sgc<\sgx<\sgm$ has to hold, we see that we need two types of
mild tunings in order to obtain successful FHI. To quantify the
amount of these tunings, we define the quantities
\beq
\Dex={\sgx-\sgc\over\sgc}~~\mbox{and}~~\Dmax=
{\sgm-\sgx\over\sgm}\,.\label{dms}\eeq
The naturalness of the hilltop FHI increases with $\Dex$ and
$\Dmax$. To get an impression of the amount of these tunings and
their dependence on the parameters of the model, we display in
\Tref{tab} the resulting $\Dex$ and $\Dmax$ together with $M$,
$\aS$, $\as$, and $r$ for $\kp=0.0005$ and $\ns$ fixed to its
central value in \Eref{nswmap}. In all cases, we obtain
$\Nhi\simeq40.5$ from \Eref{Nhi}. We notice that $\Dmax>\Dex$ and
that their values may be up to $10\%$ increasing with $\Nr$ (and
$\aS$). Recall that in \cref{mfhi} it is shown that $\Dex$ and
$\Dmax$ increase with $\kp$ (and $M$). From the observables listed
in \Tref{tab} we also infer that $|\as|$ turns out to be of order
$10^{-4}$, whereas $r$ is extremely tiny, of order $10^{-11}$, and
therefore far outside the reach of the forthcoming experiments
devoted to detect primordial gravity waves. For the preferred
$\ns$ values, we observe that $r$ and $|\as|$ increase with $\aS$.

The structure of $\Vhi$ described above is visualized in
\Fref{fig0}, where we display a typical variation of $\Vhi$ as a
function of $\sg/M$ for the values of the parameters shown in
column B of \Tref{tab}. The maximum of $\Vhi$ is located at
$\sgm/M=1.52\,\{1.38\}$, whereas its minimum lies at $\sg_{\rm
min}/M=29.1\,\{29.5\}$ -- the values obtained via the approximate
\eqs{sgm}{sgmn} are indicated in curly brackets. The values of
$\sgx/M\simeq1.4637$ and $\sgf/M\simeq1.41421$ are also depicted
together with $\sgm/M$ in the subplot of this figure. We remark
that the key $\sg$ values for the realization of FHI are squeezed
very close to one another and so their accurate determination is
essential for obtaining reliable predictions from \eqs{nS}{aS}.
Moreover, $\Nhi$ in \Eref{Nhi} can only be found numerically
taking all the possible contributions to $\Vhi'$ from
\eqs{cdev1}{cdev2} and thus $\sgx$ can not be expressed
analytically in terms of $\Nhi$. For these reasons, the results
presented in the following are exclusively based on our numerical
analysis.

\begin{figure}[!t]
\includegraphics[width=60mm,angle=-90]{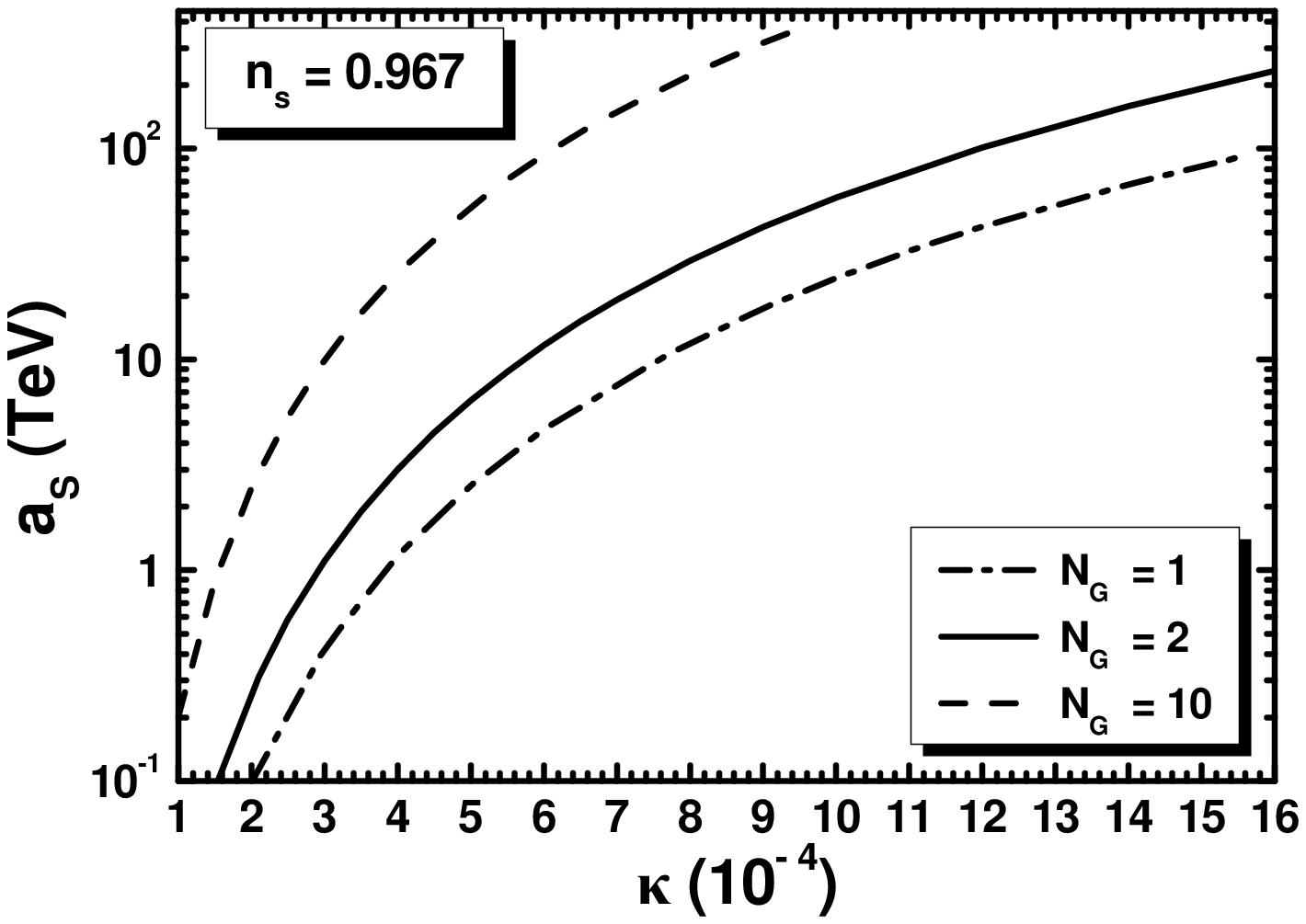}
\vspace*{-0.5cm} \caption{\sl Values of $\aS$ allowed by
Eqs.~(\ref{Nhi}) and (\ref{Prob}) versus $\kp$ for various $\Nr$'s
and fixed $\ns=0.967$.}\label{fig1}
\end{figure}

We first display in \Fref{fig1} the contours which are allowed by
Eqs.~(\ref{Nhi}) and (\ref{Prob}) in the $\kappa-\aS$ plane taking
$\ns=0.967$ and $\Nr=1$ (dot-dashed line), $\Nr=2$ (solid line),
and $\Nr=10$ (dashed line). The various lines terminate at $\kp$
values close to $10^{-3}$ beyond which no observationally
acceptable inflationary solutions are possible. We do not depict
the very narrow strip obtained for each $\Nr$ by varying $\ns$ in
its allowed range in \Eref{nswmap}, since the obtained boundaries
are almost indistinguishable. From the plotted curves we notice
that the required $\aS$'s increase with $\Nr$.

\begin{figure}[!t]
\centering
\includegraphics[width=60mm,angle=-90]{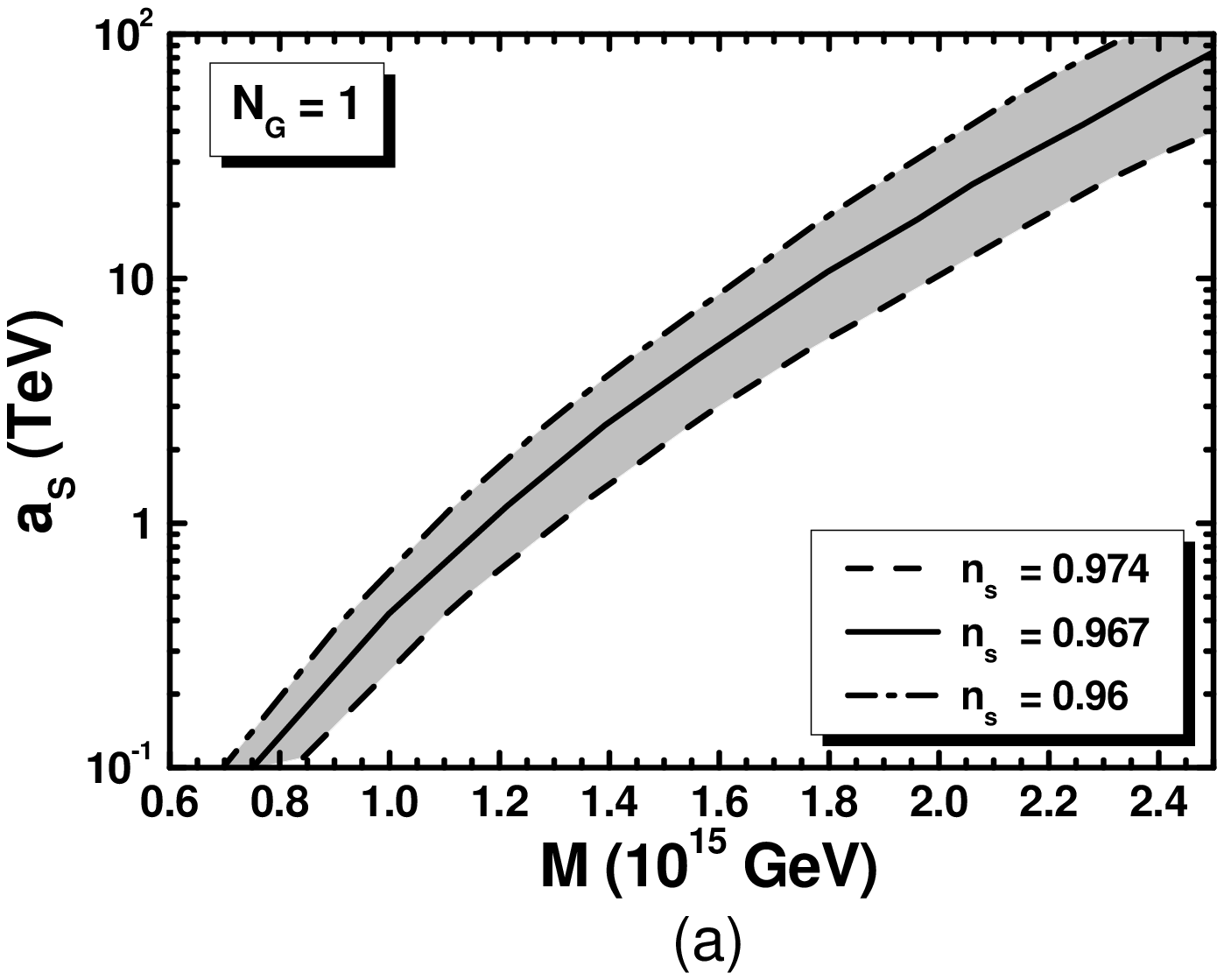}\\
\includegraphics[width=60mm,angle=-90]{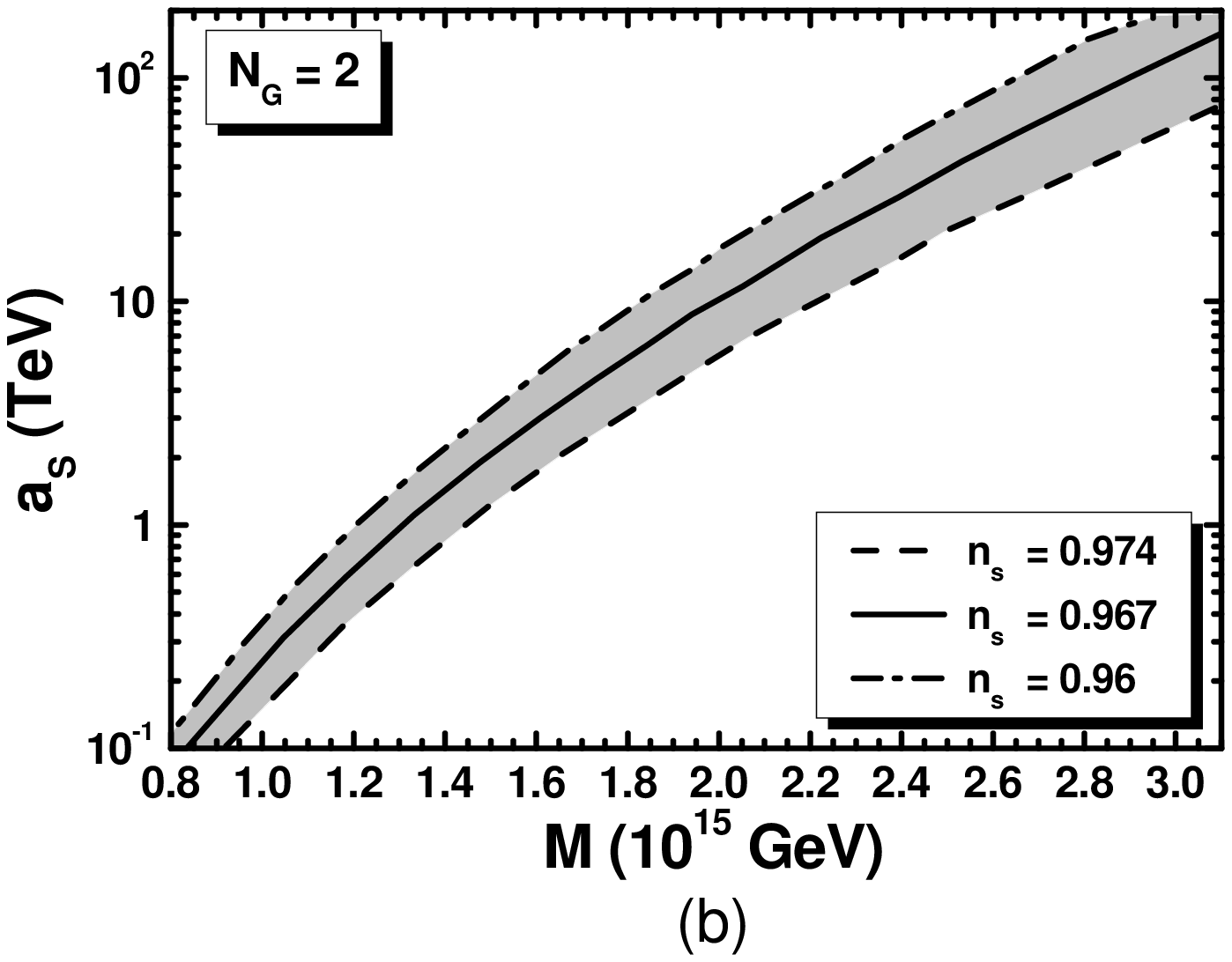}\\
\includegraphics[width=60mm,angle=-90]{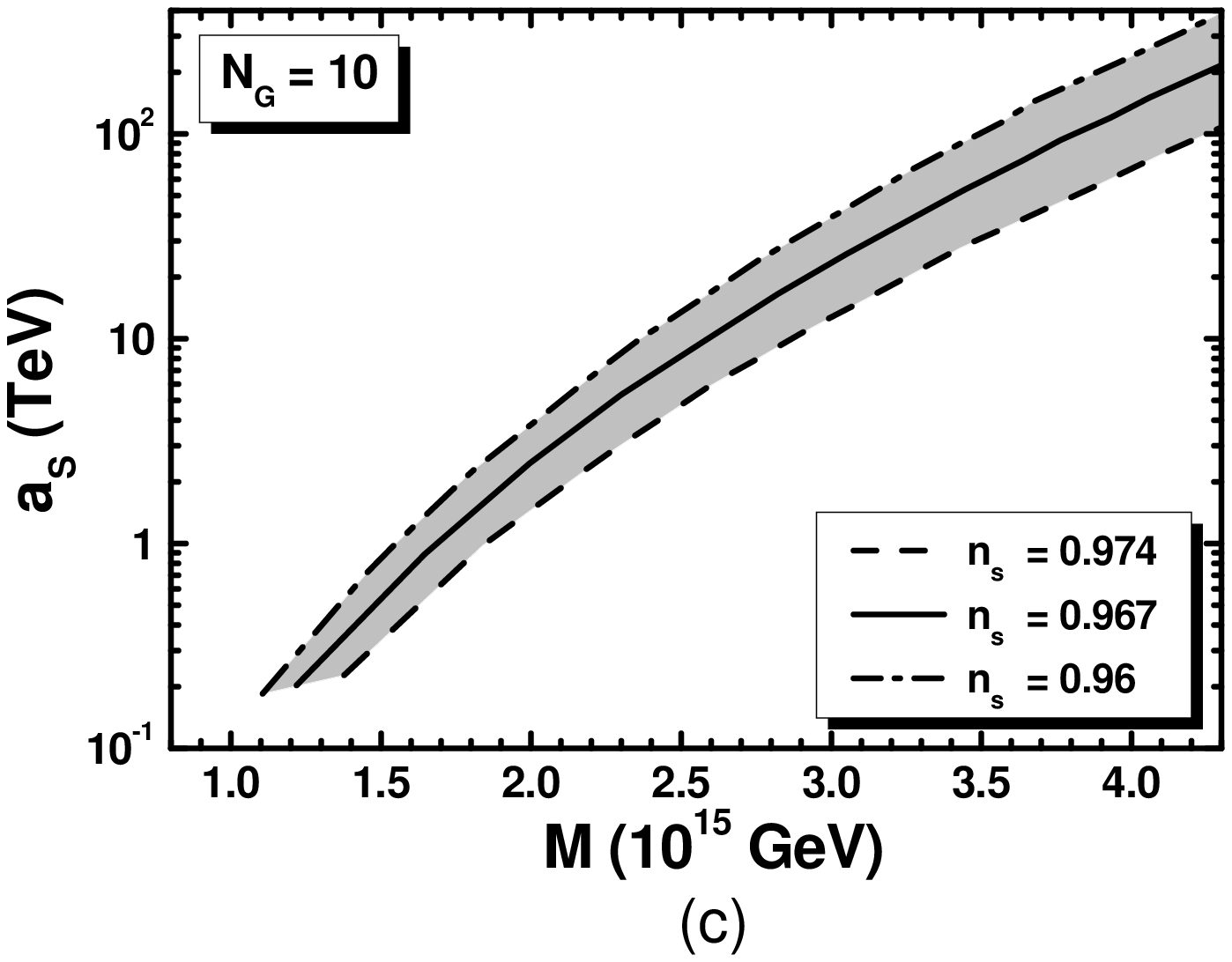}
\caption{\sl  Regions (shaded) allowed by Eqs.~(\ref{Nhi}),
(\ref{Prob}), and (\ref{nswmap}) in the $M-\aS$ plane for $\Nr=1$
(a), $\Nr=2$ (b) and $\Nr=10$ (c). The conventions adopted for the
various lines are also shown.}\label{fig2}
\end{figure}

Working in the same direction, we delineate in \Fref{fig2} the
regions in the $M-\aS$ plane allowed by Eqs.~(\ref{Nhi}),
(\ref{Prob}), and (\ref{nswmap}) for the considered $\Ggut$'s. In
particular, we use $\Nr=1,2,$ and $10$ in subfigures (a), (b), and
(c) respectively. The boundaries of the allowed areas in
\Fref{fig2} are determined by the dashed [dot-dashed] lines
corresponding to the upper [lower] bound on $n_{\rm s}$ in
Eq.~(\ref{nswmap}). We also display by solid lines the allowed
contours for $\ns=0.967$. We observe that the maximal allowed
$M$'s increase with $\Nr$. The maximal $r$'s are encountered in
the upper right end of the dashed lines, which correspond to
$\ns=0.974$, with the maximal value being $r=6.2\times10^{-10}$
for $\Nr=10$. On the other hand, the maximal $|\as|$'s are
achieved along the dot-dashed lines and the minimal value of $\as$
is $-3.2\times10^{-4}$ for $\Nr=10$ too. Summarizing our findings
from Fig.~\ref{fig2} for the central $n_{\rm s}$ value in
\Eref{nswmap} and $\Nr=1, 2,$ and $10$ respectively we end up with
the following ranges:
\beqs\bel &0.07\lesssim
{M/10^{15}~{\rm GeV}}\lesssim2.56~~\mbox{and}~~0.1\lesssim{\aS/\TeV}\lesssim100, \label{res1}  \\
&0.82\lesssim{M/10^{15}~{\rm
GeV}}\lesssim3.7~~\mbox{and}~~0.09\lesssim{\aS/\TeV}\lesssim234,
\label{res2}\\& 1.22\lesssim {M/10^{15}~{\rm
GeV}}\lesssim4.77~~\mbox{and}~~0.2\lesssim{\aS/\TeV}\lesssim460.\label{res3}\end{align}
\eeqs\\ [-0.4cm]
Within these margins, $\Dex$ ranges between $0.5\%$ and $20\%$ and
$\Dmax$ between $0.4\%$ and $12\%$. The lower bounds of these
inequalities  are expected to be displaced to slightly larger
values due to the post-inflationary requirements in
\eqs{tns1}{tns2} which are not considered here for the shake of
generality. Recall that precise incorporation of these constraints
requires the adoption of a specific $K$ from Eqs.~(\ref{K1}) --
(\ref{K4}) and corresponding $\mu/\mss$ relation from
\Eref{mueff}.

In the case $\Ggut=\Gbl$, CSs may be produced after FHI with
$\mcs=(6.5-89)\times10^{-9}$ for the parameters in \Eref{res1}.
Therefore, the corresponding parameter space is totally allowed by
Eq.~(\ref{plcs}) but completely excluded by \Eref{ppta}, if the
CSs are stable. If these CSs are metastable, the explanation
\cite{kainano} of the recent data \cite{nano, pta} on stochastic
gravity waves is possible for $M\gtrsim9\times10^{14}~\GeV$ in
\Eref{res1} where \Eref{kai} is fulfilled. No similar restrictions
exist if $\Ggut=\Glr$ or $\Gfl$, which do not lead to the
production of any cosmic defect. On the other hand, the
unification of gauge coupling constants within MSSM close to
$\Mgut=2.86\times10^{16}~\GeV$ remains intact if $\Ggut=\Gbl$
despite the fact that $M\ll\Mgut$ for $M$ given in \Eref{res1}.
Indeed, the gauge boson associated with the $U(1)_{B-L}$ breaking
is neutral under $\Gsm$ and so it does not contribute to the
relevant renormalization group running. If $\Ggut=\Glr$ or $\Gfl$
we may invoke threshold corrections or additional matter
supermultiples to restore the gauge coupling unification -- for
$\Ggut=\Gfl$ see \cref{flipnew}.



\subsection{Link to the MSSM}\label{fhi4b}

The inclusion of the HS in our numerical computation assists us to
gain information about the mass scale of the SUSY particles
through the determination of $\mss\sim\mgr$ -- see \Eref{mssi}.
Indeed, $\aS$, which is already restricted as a function of $\kp$
or $M$ for given $\Nr$  in Figs.~\ref{fig1} and \ref{fig2}, can be
correlated to $m$ via \Eref{aSn}. Taking into account \Eref{veviz}
and the fact that $\vevi{z}/\mP\sim10^{-3}$ -- see \Tref{tab} --
we can solve analytically and very accurately \Eref{aSn} w.r.t.
$m$. We find
\beq
m\simeq\lf\frac{\aS}{2^{1+\nu}(2-\nu)}\rg^{(2-\nu)/2}\lf\frac{3\Hhi^2}{(1-\nu)\nu^2}\rg^{\nu/4}.\label{maS}\eeq
Let us clarify here that in our numerical computation we use an
iterative process, which though converges quickly, in order to
extract consistently $m$ as a function of $\kp$ and $M$. This is
because the determination of the latter parameters via the
conditions in \eqs{Nhi}{Prob} requires the introduction of a trial
$m$ value which allows us to use as input the form of $\Vhi$ in
\Eref{vol}. Thanks to the aforementioned smallness of $\vevi{z}$
in \Eref{aSn}, $m$ turns out to be two to three orders of
magnitude larger than $\aS$, suggesting that $\mss$ lies clearly
at the $\PeV$ scale via \eqs{mssi}{mgr}. In fact, taking advantage
of the resulting $m$ for fixed $\nu$ in \Eref{maS}, we can compute
$\mgr$ from \Eref{mgr}, and $\mz$ and $\mth$ from \Eref{mzth}. All
these masses turn out to be of the same order of magnitude -- see
\Tref{tab}. Then $\mss$ and $\Trh$ can be also estimated from
\eqs{mssi}{Trh} for a specific $K$ from Eqs.~(\ref{K1}) --
(\ref{K4}). The magnitude of $\mss$ and the necessity for
$\mu\sim\mss$, established in \Sref{fhi}, hints towards the
high-scale MSSM.

\begin{figure}[!t]
\includegraphics[width=60mm,angle=-90]{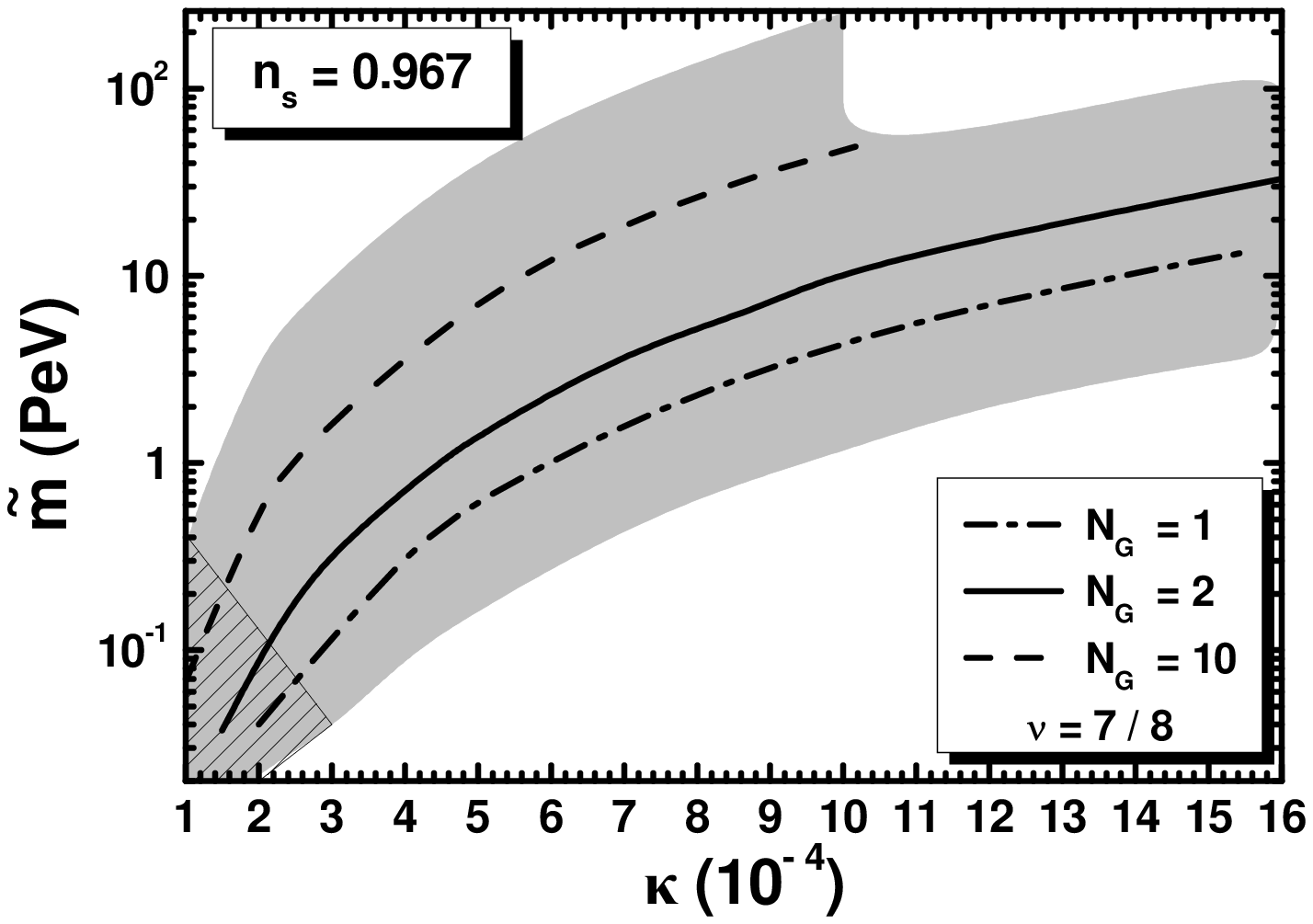}
\caption{\sl Region in the $\kp-\mss$ plane allowed by
Eqs.~(\ref{Nhi}) and (\ref{Prob})for $K=K_1$, $\mu=\mss$,
$\ns=0.967$, and $1\leq\Nr\leq10$, $3/4<\nu<1$. The allowed
contours for $\nu=7/8$ are also depicted. Hatched is the region
excluded by BBN for $\br=0.001$.}\label{fig4}
\end{figure}

To highlight numerically our expectations, we take $K=K_1$ and fix
initially $\nu=7/8$, which is a representative value. The
predicted $\mss$ as a function of $\kp$ is depicted in \Fref{fig4}
for the three $\Nr$'s considered in our work. We use the same type
of lines as in \Fref{fig1}. Assuming also that $\mu=\mss$ we can
determine the segments of these lines that can be excluded by the
BBN bound in \Eref{tns2}. In all, we find that $\mss$ turns out to
be confined in the ranges
\beqs\bea & 0.34\lesssim\mss/\PeV\lesssim 13.6
~~\mbox{for}~~\Nr=1,\label{mss1}\\
&0.21\lesssim\mss/\PeV\lesssim 32.9 ~~\mbox{for}~~\Nr=2,\label{mss2}\\
&0.58\lesssim\mss/\PeV\lesssim 46.8 ~~\mbox{for}~~\Nr=10.
\label{mss10}\eea\eeqs
Allowing $\nu$ and $\mu$ to vary within their possible respective
margins $(0.75-1)$ and $(1-3)\mss$, we obtain the gray shaded
region in \Fref{fig4}. We present an overall region for the three
possible $\Nr$'s, since the separate ones overlap each other.
Obviously the lower boundary curve of the displayed region is
obtained for $\Nr=1$ and $\nu\simeq0.751$, whereas the upper one
corresponds to $\Nr=10$ and $\nu\simeq0.99$. The hatched region is
ruled out by \Eref{tns2}. All in all, we obtain the predictions
\beq 1.2\lesssim \aS/\TeV\lesssim
460~~\mbox{and}~~0.09\lesssim\mss/\PeV\lesssim 253
\label{msst}\eeq
and $T_{\rm rh}^{\max}\simeq71~\GeV, 139~\GeV,$ and $163~\GeV$ for
$\Nr=1,2,$ and $10$ respectively attained for $\mu=3\mss$ and
$\nu\simeq0.99$. The derived allowed margin of $\mss$, which is
included in \Eref{highb}, and the employed $\mu$ values render our
proposal compatible with the mass of the Higgs boson discovered in
LHC \cite{lhc} if we adopt as a low energy effective theory the
high-scale version of MSSM \cite{strumia}.

\section{Conclusions}\label{con}

We considered the realization of FHI in the context of an extended
model based on the superpotential and \Kaa\ potential in
\eqs{Who}{Kho}, which are consistent with an approximate $R$
symmetry. The minimization of the SUGRA scalar potential at the
present vacuum constrains the curvature of the internal space of
the goldstino superfield and provides a tunable energy density
which may be interpreted as the DE without the need of an
unnaturally small coupling constant. On the other hand, this same
potential causes a displacement of the sgoldstino to values much
smaller than $\mP$ during FHI. Combining this fact with minimal
kinetic terms for the inflaton, the $\eta$ problem is resolved
allowing hilltop FHI. The slope of the inflationary path is
generated by the RCs and a tadpole term with a minus sign and
values which increase with the dimensionality of the
representation of the relevant Higgs superfields. Embedding $\Gbl$
into a larger gauge group $\Gu$ which predicts the production of
monopoles prior to FHI that can eventually break the CSs allows
the attribution of the observed data on the gravitational waves to
the decay of metastable $B-L$ CSs.

We also discussed the generation of the $\mu$ term of MSSM
following the Giudice-Masiero mechanism and restricted further the
curvature of the goldstino internal space so that
phenomenologically dangerous production of $\Gr$ may be avoided.
This same term assists in the decay of the sgoldstino, which
normally dominates the energy density of the Universe, at a reheat
temperature which can be as high as $163~\GeV$ provided that the
$\mu$ parameter is of the order of the $\Gr$ mass, i.e., of order
$\PeV$. Linking the inflationary sector to a degenerate MSSM mass
scale $\mss$ we found that $\mss$ lies in a range consistent with
the Higgs boson mass measured at LHC within high-scale SUSY.

The long-lasting matter domination obtained in our model because
of the sgoldstino oscillations after the end of FHI leads
\cite{wells} to a suppression at relatively large frequencies
($f>0.1~{\rm Hz}$) of the spectrum of the gravitational waves from
the decay of the metastable CSs. This effect may be beneficial for
spectra based on $\mcs$ values which violate the upper bound of
\Eref{kai} from the results of \cref{ligo}. Since we do not
achieve such $\mcs$ values here we do not analyze further this
implication of our scenario. On the other hand, the low reheat
temperature encountered in our proposal makes difficult the
achievement of baryogenesis. However, there are currently attempts
\cite{bau} based on the idea of cold electroweak baryogenesis
\cite{cbau} which may overcome this problem. It is also not clear
which particle could play the role of CDM in a high-scale SUSY
regime. Let us just mention that a thorough investigation is
needed including the precise solution of the relevant Bolzmann
equations as in \cref{rh} in order to assess if the abundance of
the lightest SUSY particle can be confined within the
observational limits in this low-reheating scenario.


\acknowledgments  We would like to thank I. Antoniadis, H. Baer,
and E.~Kiri-tsis for useful discussions. This research work was
supported by the Hellenic Foundation for Research and Innovation
(H.F.R.I.) under the ``First Call for H.F.R.I. Research Projects
to support Faculty members and Researchers and the procurement of
high-cost research equipment grant'' (Project Number: 2251).


\appendix*

\renewenvironment{subequations}{%
\refstepcounter{equation}%
\setcounter{parentequation}{\value{equation}}%
  \setcounter{equation}{0}
  \def\theequation{A\theparentequation{\small\sffamily\alph{equation}}}%
  \ignorespaces
}{%
  \setcounter{equation}{\value{parentequation}}%
  \ignorespacesafterend
}

\newcommand{\kf}{\ensuremath{k_{4}}}
\newcommand{\ksx}{\ensuremath{k_{6}}}
\renewcommand\vevi[1]{\left\langle {#1} \right\rangle_{\rm I}}
\renewcommand\mgr{m_{\rm I3/2}}

\section{SUGRA CORRECTIONS TO THE INFLATIONARY POTENTIAL of FHI}

As shown in \Sref{asfhi3a}, the presence of $\whi$ and $K_{\rm H}$
in \eqs{wh}{khi} respectively transmit (potentially important)
corrections to the inflationary potential. We present here, for
the first time to the best of our knowledge, these corrections
without specify the form of these functions. The corrections from
the IS are also taken into account.

In particular, we consider the following superpotential and \Kaa\
potential resulting from the ones in \eqs{Who}{Kho} by setting
$\phc$ and $\phcb$ to zero:
\beq   W=W_{\rm I}(S) +W_{\rm H}(Z) ~~\mbox{and}~~ K=K_{\rm
I}(S)+K_{\rm H}(Z), \label{wkgen}\eeq
where $\Whi$ and $\khi$ are given by
\beq \Whi=-\hkp M^2 S~~\mbox{and}~~
\khi=\khi(|S|^2)\label{whi0}\eeq
(cf. \eqs{whi}{ki}). We also assume that $\khi$ can be reliably
expanded in powers of $|S|/\mP$ as follows:
\beq K_{\rm I}\simeq|S|^2+\frac{\kf}{4}
\frac{|S|^4}{\mP^2}+\frac{\ksx}{9}
\frac{|S|^6}{\mP^3}+\cdots.\label{ksg}\eeq
Under these circumstances, the inverse K\"ahler metric reads
\beqs\beq K^{SS^*}_{\rm I} \simeq1-\kf
{|S|^2}{\mP^2}+(\kf^2-\ksx)
{|S|^4}/{\mP^4}+\cdots\label{kimsg}\eeq
and the exponential prefactor of $\vf$ in \Eref{Vsugra} is well
approximated by
\beq e^{K_{\rm I}/\mP^2}\simeq
1+\frac{|S|^2}{\mP^2}+\frac{1+2\kf}{2}
\frac{|S|^4}{\mP^4}+\cdots.\label{eexp}\eeq\eeqs

Taking into account the two last expressions and expanding $\vf$
in \Eref{Vsugra} with $W$ and $K$ from \Eref{wkgen} up to the
forth power in $|S|/\mP$, we obtain the quite generic formula
below
\bel \vf&\simeq v_{0}+\mgr^2|S|^2+\lf v_1 S^* +{\rm
c.c}\rg+v_2{|S|^2}/{\mP^2}\nonumber\\& + \lf v_3 S^* +{\rm
c.c}\rg{|S|^2}/{\mP^2}+v_4{|S|^4}/{\mP^4}+\cdots, \label{vsgg}
\end{align}
where the various $v$'s are found to be
\beqs\bel
v_0&=\kp^2 M^4,\label{vsg0g}\\
v_1&=\kp M^2\mgr\vevi{2-K^{ZZ^*}_{\rm H}\partial_Z G_{\rm H}},\label{vsg1g}\\
v_2&=\kp^2 M^4 \vevi{K^{ZZ^*}_{\rm H}|\partial_Z K_{\rm H}|^2/\mP^2-\kf},\label{vsg2g}\\
v_3&=\kp M^2\mgr\vevi{(1+\kf/2)-K^{ZZ^*}_{\rm H}\partial_Z G_{\rm H}},\label{vsg3g}\\
v_4&= \kp^2 M^4\Big(1/2+ \kf(4\kf-7)/4-\ksx\nonumber\\
&\left. +\vevi{K^{ZZ^*}_{\rm H}|\partial_Z K_{\rm H}|^2/\mP^{2}}
\rg.\label{vsg4g}
\end{align}\eeqs
%
Here $\kp$ is the rescaled coupling constant $\hkp$ after
absorbing the relevant prefactor $e^{\vevi{K_{\rm H}}/2\mP}$ in
\Eref{Vsugra} and we used the definition of the $\Gr$ mass
$$\mgr=\vevi{e^{K_{\rm H}/2\mP^2}\whi/\mP^2},$$ and the
K\"{a}hler invariant function, see, e.g., \cref{gref},
\beq G_{\rm H}=K_{\rm H}/\mP^2+\ln|W_{\rm H}/\mP^3|^2.\eeq

From these results we see that $v_2$ and $v_4$ generically receive
contributions from both the IS and HS, whereas $v_1$ and $v_3$
exclusively from the HS -- cf. Refs.~\cite{pana,kelar}.
Specifically, from \Eref{vsg2g}, we can recover the miraculous
cancellation occurring within minimal FHI \cite{mfhi,buchbl},
where the HS is ignored and $k_4=k_6=0$ in \Eref{ksg}. Switching
on $K_{\rm H}$ and noticing that \beq
\kf=\partial^2_S\partial^2_{S^*}
\khi(S=S^*=0),\label{kssbpana}\eeq we can also see that
\Eref{vsg2g} agrees with that presented in \cref{pana}. The
applicability of our results can be easily checked for other HS
settings \cite{buch1,high,nshi} too.

\def\ijmp#1#2#3{{\sl Int. Jour. Mod. Phys.}
{\bf #1},~#3~(#2)}
\def\plb#1#2#3{{\sl Phys. Lett. B }{\bf #1}, #3 (#2)}
\def\prl#1#2#3{{\sl Phys. Rev. Lett.}
{\bf #1},~#3~(#2)}
\def\rmp#1#2#3{{Rev. Mod. Phys.}
{\bf #1},~#3~(#2)}
\def\prep#1#2#3{{\sl Phys. Rep. }{\bf #1}, #3 (#2)}
\def\prd#1#2#3{{\sl Phys. Rev. D }{\bf #1}, #3 (#2)}
\def\npb#1#2#3{{\sl Nucl. Phys. }{\bf B#1}, #3 (#2)}
\def\npps#1#2#3{{Nucl. Phys. B (Proc. Sup.)}
{\bf #1},~#3~(#2)}
\def\mpl#1#2#3{{Mod. Phys. Lett.}
{\bf #1},~#3~(#2)}
\def\jetp#1#2#3{{JETP Lett. }{\bf #1}, #3 (#2)}
\def\app#1#2#3{{Acta Phys. Polon.}
{\bf #1},~#3~(#2)}
\def\ptp#1#2#3{{Prog. Theor. Phys.}
{\bf #1},~#3~(#2)}
\def\n#1#2#3{{Nature }{\bf #1},~#3~(#2)}
\def\apj#1#2#3{{Astrophys. J.}
{\bf #1},~#3~(#2)}
\def\mnras#1#2#3{{MNRAS }{\bf #1},~#3~(#2)}
\def\grg#1#2#3{{Gen. Rel. Grav.}
{\bf #1},~#3~(#2)}
\def\s#1#2#3{{Science }{\bf #1},~#3~(#2)}
\def\ibid#1#2#3{{\it ibid. }{\bf #1},~#3~(#2)}
\def\cpc#1#2#3{{Comput. Phys. Commun.}
{\bf #1},~#3~(#2)}
\def\astp#1#2#3{{Astropart. Phys.}
{\bf #1},~#3~(#2)}
\def\epjc#1#2#3{{Eur. Phys. J. C}
{\bf #1},~#3~(#2)}
\def\jhep#1#2#3{{\sl J. High Energy Phys.}
{\bf #1}, #3 (#2)}
\newcommand\jcap[3]{{\sl J.\ Cosmol.\ Astropart.\ Phys.\ }{\bf #1}, #3 (#2)}
\newcommand\njp[3]{{\sl New.\ J.\ Phys.\ }{\bf #1}, #3 (#2)}
\def\prdn#1#2#3#4{{\sl Phys. Rev. D }{\bf #1}, no. #4, #3 (#2)}
\def\jcapn#1#2#3#4{{\sl J. Cosmol. Astropart.
Phys. }{\bf #1}, no. #4, #3 (#2)}
\def\epjcn#1#2#3#4{{\sl Eur. Phys. J. C }{\bf #1}, no. #4, #3 (#2)}

\end{document}